\documentclass[11pt]{article}

\usepackage[margin=1in]{geometry}
\usepackage{graphicx}
\usepackage[numbers,compress,sort]{natbib}
\usepackage{hyperref}
\usepackage{caption}
\usepackage{bm}
\usepackage{xcolor}
\usepackage{placeins}
\usepackage{tikz}
\usepackage{amsmath}
\usepackage{amssymb}
\usepackage{subcaption}
\usepackage{setspace}
\usepackage{authblk}

\usetikzlibrary{arrows.meta,positioning,shapes.geometric,calc,shapes.symbols}
\usetikzlibrary{decorations.pathreplacing,shadows.blur}
\tikzset{
  pageSide/.style={
    draw, thick, rounded corners=4pt,
    inner sep=0pt,
    minimum width=1.55cm,
    minimum height=1.15cm,
    transform shape,
    blur shadow={shadow blur steps=5, shadow xshift=0.8pt, shadow yshift=-0.8pt}
  }
}
\usetikzlibrary{calc,fit,backgrounds}

\makeatletter
\pgfdeclareshape{preparation}{
  \inheritsavedanchors[from=rectangle]
  \inheritanchorborder[from=rectangle]
  \inheritanchor[from=rectangle]{center}
  \inheritanchor[from=rectangle]{north}
  \inheritanchor[from=rectangle]{south}
  \inheritanchor[from=rectangle]{east}
  \inheritanchor[from=rectangle]{west}
  \inheritanchor[from=rectangle]{north east}
  \inheritanchor[from=rectangle]{north west}
  \inheritanchor[from=rectangle]{south east}
  \inheritanchor[from=rectangle]{south west}

  \backgroundpath{
    \southwest \pgf@xa=\pgf@x \pgf@ya=\pgf@y
    \northeast \pgf@xb=\pgf@x \pgf@yb=\pgf@y

    \pgf@yc=.5\pgf@ya \advance\pgf@yc by .5\pgf@yb

    \pgfmathsetlength{\pgf@xc}{0.12* (\pgf@xb-\pgf@xa)}

    \pgfpathmoveto{\pgfpoint{\pgf@xa+\pgf@xc}{\pgf@yb}}
    \pgfpathlineto{\pgfpoint{\pgf@xb-\pgf@xc}{\pgf@yb}}
    \pgfpathlineto{\pgfpoint{\pgf@xb}{\pgf@yc}}
    \pgfpathlineto{\pgfpoint{\pgf@xb-\pgf@xc}{\pgf@ya}}
    \pgfpathlineto{\pgfpoint{\pgf@xa+\pgf@xc}{\pgf@ya}}
    \pgfpathlineto{\pgfpoint{\pgf@xa}{\pgf@yc}}
    \pgfpathclose}}
\makeatother

\definecolor{mygray}{gray}{0.65}
\definecolor{myred}{RGB}{210,60,60}
\definecolor{myblue}{RGB}{60,90,190}
\definecolor{mygreen}{RGB}{70,160,90}
\definecolor{myorange}{RGB}{230,120,20}
\definecolor{mydarkgray}{gray}{0.4} 
\tikzset{
  >={Latex[length=3mm]},
  box/.style={draw, thick, rectangle, minimum width=6.4cm, minimum height=2.2cm, align=center, inner sep=7pt},
  bigbox/.style={draw, thick, rectangle, minimum width=8.5cm, minimum height=4.2cm, inner sep=6pt},
  oval/.style={draw, thick, ellipse, minimum height=1.5cm, minimum width=0.5cm, align=center, inner sep=1pt},
  ovalSmall/.style={draw, thick, ellipse, minimum height=0.9cm, minimum width=1.5cm, align=center, inner sep=4pt},
  rrect/.style={draw=mygreen, thick, rounded corners=13pt, minimum height=1.05cm, minimum width=2.9cm, align=center, inner sep=2pt},
  polyG/.style={draw=mygreen, thick,shape=preparation,minimum width=5.5cm,minimum height=2.25cm,align=center,inner xsep=0pt,inner ysep=0pt},
}

\tikzset{bigarrow/.style={-{Stealth[length=12pt,width=12pt]},line width=2pt}}
\tikzset{bigarrow2/.style={-{Stealth[length=12pt,width=20pt]},line width=2pt}}

\hypersetup{
    colorlinks = true,
    urlcolor   = blue,
    citecolor  = black,
}

\title{\Large\bfseries Training of particle-turbulence sub-grid-scale closures\\ with just particle data}

\author{German Saltar Rivera\thanks{Corresponding author: \href{mailto:germans3@illinois.edu}{germans3@illinois.edu}}}
\author{Laura Villafa\~{n}e}
\author{Jonathan B. Freund}
\affil{\small Department of Aerospace Engineering, University of Illinois Urbana--Champaign, United States of America}
\date{}

\begin{document}
\maketitle
\vspace{-1em}

\begin{abstract}
If sufficient training data are available, neural networks are attractive for representing missing physics in simulations, such as sub-grid scales in the coarse-mesh particle-turbulence system we consider. Physical constraints are known to both increase performance and reduce the need for data; we use the complete physics represented in the discretized governing equations as a constraint. Two-way coupled particles in two-dimensional turbulence provide a sufficiently complex system to assess effectiveness for various training data, all constructed from well-resolved simulations, in cases intentionally degraded to assess robustness. Surprisingly, using the full space-time data actually hinders model effectiveness. Instead, training that targets only spectra---hence, neglecting phase information---provides better closures, which is related to the well-known success of non-dissipative discretizations for simulating turbulence. It is found that some of the missing physics that lead to preferential particle concentration errors are fundamentally stochastic on the coarse mesh and therefore uncorrectable by the basic approach; a learning formulation is introduced for a Langevin-type closure to correct this. Most importantly, training just for particle kinetic energy---without any direct input from the flow field---also yields effective sub-grid-scale stress models. This holds true even if noise is added to the particle data, if only a sub-sample of particles are used, or if only one component of the particle velocity is used.  In sum, these results show a path for inferring sub-grid-scale physics based just on particle data from experiments. 
\end{abstract}

\section{Introduction}\label{sec:intro}
Turbulent particle-laden flows are central to phenomena ranging from atmospheric dust transport to combustion in engines.
Accurate prediction of these flows is therefore important, especially when experimental data are scarce, expensive, or difficult to obtain.
To reduce the full computational costs of representing all the turbulence scales, many simulation approaches represent only the large, energy-containing turbulence scales while modeling the effects of the unresolved scales, which are relatively universal.

We consider such large-eddy simulation (LES) in cases for which the particle diameter $d_p$ is much smaller than the Kolmogorov length scale $\eta$ and so particles are represented as points~\cite{balachandar_turbulent_2010,elghobashi_predicting_1994,kuerten_collision_2016}.
We also consider cases with particle volume fraction $\Phi_v = N_p V_p/V$ in the range $10^{-6}<\Phi_v<10^{-3}$, for which two-way coupling significantly affects the flow and particle kinematics~\cite{elghobashi_predicting_1994}.
In this regime, unrepresented small turbulence scales affect particle concentrations and accelerations~\cite{armenio_effect_1999,cernick_particle_2015,he_space-time_2017,jin_large-eddy_2010,kuerten_can_2005,marchioli_issues_2008,park_simple_2017,ray_preferential_2011,urzay_characteristic_2014}.
There are many approaches to modeling particle-laden turbulence in this limit~\cite{berrouk_stochastic_2007,bini_particle_2007,fede_stochastic_2006,fukagata_dynamics_2004,gorokhovski_lagrangian_2014,gosman_aspects_1983,jin_nonlinear_2013,mallouppas_large_2013,minier_lagrangian_2015,pozorski_filtered_2009,shotorban_stochastic_2006,kuerten_subgrid_2006,park_simple_2017,shotorban_modeling_2005,gobert_subgrid_2011,murray_single-particle_2016,ray_subgrid_2014}; the machine-learning (ML) approach we consider provides a means to improve them and a path for incorporating additional sub-grid-scale physics. 
Results also provide a route for how sub-grid-scale (SGS) models might be deduced from quantities far more easily measured in experiments.

Neural networks have a remarkable capacity to fit the complex data~\cite{brenner_perspective_2019,brunton_machine_2020,choi2025perspective}.
The most common approach to bring them into modeling is a straightforward supervised ML procedure where a needed closure $\mathbf{h}$ is learned directly (\textit{a priori}) from trusted $\bm{h}_\mathrm{known}$ sub-grid-scale data (such as stress tensor $\bm{\tau}^r$ or the sub-grid-scale velocity perturbations $\mathbf{u}^\prime$)~\cite{beck_deep_2019,cho_recursive_2023,gamahara_searching_2017,hu_improving_2024,kim_large_2024,maulik_subgrid_2019,ren_artificial-neural-network-based_2025,wu_large-eddy_2022}.
For combined flow and particle state $\mathbf{q} = \mathbf{q}(\mathbf{u},\mathbf{x}_\mathrm{p},\mathbf{v}_\mathrm{p})$, the neural-network training is straightforward:
\begin{equation}\label{eq:apriori_opt}
	\begin{split}
		\text{find neural-network weights}\quad &\vec{\theta},\\
		\text{such that SGS closure} \quad & \mathbf{h}(\bm{q};\vec{\theta}\,)\approx \bm{h}_\mathrm{known}.
	\end{split}
\end{equation} 

Of course, this strategy faces a challenge for sub-grid-scale closures: it requires space- and time-resolved data from high-resolution simulations or experimental diagnostics, which are expensive and difficult to obtain.
For the discussion here, it is important to note that, for large-eddy simulation, reproducing the instantaneous sub-grid-scale data does not necessarily align with the actual goals of a closure. 
\cite{langford_optimal_1999} provide a pithy exposition of this: they show that a substantial part of $\bm{\tau}^r$ is effectively stochastic in the sense that it is uncorrelated with the represented flow quantities. 
Hence, no deterministic closure is able to predict the true instantaneous sub-grid-scale quantities based on the represented field.
We show in section~\ref{sec:full_obj} that training neural networks against actual values in this way indeed over-constrains them and hinders predictions.\@
The disconnect created by information lost to filtering is only expected to be accentuated with additional physics, such as the two-way coupled particle dynamics we consider.

This disconnect motivates an approach that trains on the actual objective: relevant particle and turbulence dynamics, thus avoiding the consequence of the lost information that renders true sub-grid scales effectively noise.\@
The needed closure (the learned sub-grid-scale model) for the prediction objective (the outcome of the large eddy simulation) can be obtained through a more involved optimization problem:
\begin{equation}\label{eq:apost_opt}
	\begin{split}
		\text{find neural-network weights}\quad &\vec{\theta},\\
		\text{such that SGS closure} \quad & \mathbf{h}(\bm{q};\vec{\theta}\,),\\
		\text{best predicts state} \quad & \mathbf{q} \;\;[\text{or statistics}\;\: S(\mathbf{q})], \\
		\text{subject to governing equation} \quad & \mathcal{N}(\mathbf{q})=0.
	\end{split}
\end{equation} 
This approach has been successfully demonstrated for sub-grid-scale models in single-phase incompressible turbulence~\cite{sirignano_dpm_2020}, scalar transport~\cite{macart_embedded_2021}, flow over bluff bodies~\cite{sirignano_deep_2023}, and for Reynolds-averaged simulation of flow around airfoils~\cite{holland_field_2019}.
For particle-laden turbulence, the effects of sub-grid-scale flow--particle interactions must also be accounted for.
We analyze this challenge and show that this approach is sufficient to develop a working sub-grid-scale closure for turbulence using only the particle kinetic energy distribution, without any flow information. This extension requires significant development and reveals some unexpected factors to consider.
Unlike physics-informed neural networks (PINNs), which enforce physics by minimizing pointwise PDE residuals at sampled locations and times~\cite{coscia_physics-informed_2023,karniadakis_physics-informed_2021,krishnapriyan2021pinn_failure_modes,raissi_physics-informed_2019}, the optimization problem in (\ref{eq:apost_opt}) enforces the governing equations as constraints through the large-eddy simulation solver. 
Consequently, closures are updated based on their effect on the actual simulation outcomes.

The central challenge to doing this is obtaining the gradient needed to efficiently learn optimal weights $\vec{\theta}$.
To obtain the sensitivity of the loss function $\mathcal{J}(\mathbf{q})$ to $\vec{\theta}$, which in our formulation depends on the solution $\mathbf{q}$, discrete-exact adjoints are used to compute the $\frac{\partial\mathbf{q}}{\partial\mathbf{h}}$ component of
\begin{equation}
	\frac{\partial \mathcal{J}}{\partial \vec{\theta}} = \frac{\partial \mathcal{J}}{\partial\mathbf{q}}\cdot\frac{\partial\mathbf{q}}{\partial\mathbf{h}}\cdot\frac{\partial\mathbf{h}}{\partial\vec{\theta}}\, ,
\end{equation}
which is used to update $\vec{\theta}$.
This is formulated in section~\ref{sec:PCO}.

We focus on two-dimensional turbulence as a low-cost, well-understood testbed to develop and demonstrate our approach~\cite{heijst_laboratory_2009,kellay_two-dimensional_2002,kraichnan_inertial_1967,kraichnan_two-dimensional_1980,tabeling_experimental_1991}.
In contrast to three-dimensional turbulence, it is a dual-cascade system with inverse energy and forward enstrophy cascades, yet it presents a multiple-scales challenge analogous to three-dimensional turbulence.
Key coupling mechanisms between particles and turbulence are present, including preferential concentration and turbulence modulation~\cite{boffetta_large_2004,onishi_collision_2014,pandey_clustering_2019}.

The following section~\ref{sec:formulation} introduces the governing equations, followed by their use as training constraints (section~\ref{sec:PCO}) and the high-resolution simulations used to generate training data (section~\ref{sec:training_data}).
A generic neural-network architecture, implemented in a way that conserves momentum, is described along with the training algorithm in section~\ref{sec:NN_design}.
Results analyzed in section~\ref{sec:demonstration} illustrate challenges and lead to an effective closure that can be trained on limited and noisy data.

\section{Formulation}\label{sec:formulation}
\subsection{Governing equations}
The carrier phase is incompressible with the common hyperviscosity $\propto\nabla^4\mathbf{u}$ and hypofriction $\propto\nabla^{-8}\mathbf{u}$ terms that are usually added in two-dimensional turbulence simulations to allow for a range of inertial scales~\cite{boffetta_two-dimensional_2012}.
A coupling term $\mathbf{F}_\mathrm{p}$ accounts for the momentum exchange with the particles. 

The forcing $\widehat{\mathbf{F}}_\omega$ that sustains the turbulence is similar to typical Markovian forcing~\cite{san_stationary_2013}, but without a truly stochastic temporal evolution in order to provide consistency between the DNS and LES datasets in training, as needed for parts of our analysis~\cite{babiano_vorticity_1987,kida_route_1989,ohkitani_wave_1991}.
Specifically, vorticity is forced in a narrow spectral band centered at wavenumber $\kappa_f$:
\begin{equation}\label{eq:A_omega}
	\widehat{\mathbf{F}}_\omega = 
	\begin{cases}
		ae^{i\beta}  &   \kappa_f - c \leq |\bm{\kappa}|\leq \kappa_f + c  \\
		0 & \text{otherwise},
	\end{cases}
\end{equation}
with $c=2$, $a=20$, and uniformly distributed random phases $\beta\in [0,2\pi)$. 
A divergence-free forcing
\begin{equation}\label{eq:A}
	\widehat{\mathbf{F}}_\mathrm{u} = 
	\begin{bmatrix}
		\frac{i\kappa_2}{|\bm{\kappa}|^2}\widehat{\mathrm{F}}_{\omega_1} \\
		-\frac{i\kappa_1}{|\bm{\kappa}|^2}\widehat{\mathrm{F}}_{\omega_2}
	\end{bmatrix},
\end{equation}
whose curl is $\widehat{\mathbf{F}}_\omega$, is inverse Fourier transformed and added to the flow momentum governing equation.
Reference values used for nondimensionalizing the governing equations are the fluid density $\rho^\star$, the forcing wavenumber $\kappa_f$, and the forcing amplitude $a$ (which gives a time scale $t^\star = 1/\sqrt{a}$). 

The flow governing equations are thus
\begin{subequations}\label{eq:NS}
	\begin{align}
		\nabla \cdot \mathbf{u}&=0,\\
		\frac{\partial \mathbf{u}}{\partial t} + \nabla\cdot\mathbf{u}\mathbf{u} &= - \nabla p - \upsilon\nabla^4\mathbf{u} - \alpha\nabla^{-8}\mathbf{u} -\mathbf{F}_\mathrm{p}+\mathbf{F}_\mathrm{u} ,
	\end{align}
\end{subequations}
with the flow velocity $\mathbf{u}$, pressure $p$, hyperviscosity $\upsilon$, hypofriction $\alpha$, and the particle coupling $\mathbf{F}_\mathrm{p} = m_p\frac{d\mathbf{v}_\mathrm{p}}{dt}\delta(\mathbf{x}-\mathbf{x}_\mathrm{p})$, where $\delta$ is a Dirac function, $\mathbf{v}_\mathrm{p}$ is the particle velocity, and $\mathbf{x}_\mathrm{p}$ its location.
A single relaxation time $\tau_p$ yields the usual simplified particle equations~\cite{gautignol_faxen_1983,maxey_equation_1983}:
\begin{subequations}
	\begin{align}\label{eq:cont_MR}
		\frac{d \mathbf{v}_\mathrm{p}}{dt} & =  \frac{1}{\tau_p}[\mathbf{u}(\mathbf{x}_\mathrm{p})-\mathbf{v}_\mathrm{p}], \\ \label{eq:cont_PK}
		\frac{d \mathbf{x}_\mathrm{p}}{dt} & = \mathbf{v}_\mathrm{p}.
	\end{align}
\end{subequations}
 
\subsection{Sub-grid-scale closure}
The represented scales are defined, at least in principle, based on a low-pass filter
\begin{equation}\label{eq:filter}
	\overline{\phi}(\mathbf{x},t) = \int_\Omega G(\mathbf{x}^\prime,\mathbf{x})\phi(\mathbf{x}-\mathbf{x}^\prime,t)\, d\mathbf{x}^\prime,
\end{equation}
where $G(\mathbf{x}^\prime,\mathbf{x})$ is the filter kernel.
Applying (\ref{eq:filter}) to (\ref{eq:NS}) yields
\begin{subequations}
	\begin{align}
		\nabla\cdot\overline{\mathbf{u}} &= 0,\\
		\frac{\partial \overline{\mathbf{u}}}{\partial t} + \nabla\cdot\overline{\mathbf{u}} \,\overline{\mathbf{u}} & = - \nabla p - \upsilon\nabla^4\overline{\mathbf{u}} - \alpha\nabla^{-8}\overline{\mathbf{u}} -\overline{\mathbf{F}}_\mathrm{p} +\overline{\mathbf{F}}_\mathrm{u} + \nabla\cdot\bm{\tau}^r,
	\end{align}
\end{subequations}
where $\bm{\tau}^r = \overline{\mathbf{u}}\,\overline{\mathbf{u}} - \overline{\mathbf{u}\mathbf{u}}\equiv \mathbf{h}_\mathrm{u}$ is the unclosed sub-grid-scale stress.  
The velocity field at the particle location in (\ref{eq:cont_MR}) has both represented $\overline{\mathbf{u}}(\mathbf{x}_\mathrm{p})$ and sub-grid-scale $\mathbf{u}^\prime(\mathbf{x}_\mathrm{p})$ components
\begin{align}\label{eq:MR_unclosed}
	  \frac{d \mathbf{v}_\mathrm{p}}{dt} = \frac{1}{\tau_p}[\overline{\mathbf{u}}(\mathbf{x}_\mathrm{p})-\mathbf{v}_\mathrm{p}]+\frac{1}{\tau_p}\mathbf{u}^\prime(\mathbf{x}_\mathrm{p}),
\end{align}
where $\mathbf{u}^\prime(\mathbf{x}_\mathrm{p})/\tau_p\equiv \mathbf{h}_\mathrm{p}$ is unclosed.

Neural network closures replace the unclosed terms 
\begin{subequations}\label{eq:cont_gov_NN}
	\begin{align}
			  \frac{\partial \overline{\mathbf{u}}}{\partial t} + \nabla\cdot\overline{\mathbf{u}} \,\overline{\mathbf{u}} & = - \nabla \overline{p} - \upsilon\nabla^4\overline{\mathbf{u}} - \alpha\nabla^{-8}\overline{\mathbf{u}} -\overline{\mathbf{F}}_\mathrm{p}+\overline{\mathbf{F}}_\mathrm{u}+\boxed{\nabla\cdot \mathbf{h}_\mathrm{u}(\bm{\phi}_u;\vec{\theta}_{u})},\label{eq:cont_NN_NS}\\
		 \frac{d \mathbf{v}_\mathrm{p}}{dt}  &= \frac{1}{\tau_p}[\overline{\mathbf{u}}(\mathbf{x}_\mathrm{p})-\mathbf{v}_\mathrm{p}]+\boxed{\mathbf{h}_\mathrm{p}(\bm{\phi}_p;\vec{\theta}_{p})}\label{eq:cont_NN_MR},
	\end{align}
\end{subequations}
where $\bm{\phi}_u(\mathbf{u})$ and $\bm{\phi}_p(\mathbf{u},\mathbf{v}_\mathrm{p},\mathbf{x}_\mathrm{p})$ are the neural-network inputs, which will be defined in section~\ref{sec:NN model}.

\subsection{Discretization}
Although pseudo-spectral discretizations are often preferred in periodic domains, the closure framework developed here is intended to remain applicable in more general configurations.
Accordingly, standard second-order centered finite differences on a staggered mesh and fourth-order Runge--Kutta (RK4) are used within a fractional-step method~\cite{chorin_numerical_1968}.  
For the following development, the discretizations of (\ref{eq:cont_PK}) and (\ref{eq:cont_gov_NN}) are written compactly as 
\begin{subequations}\label{eq:disc_gov}
	\begin{align}
		\vec{\mathbf{N}}^{n,s}&=0,\\
		\vec{\mathbf{M}}^{n,s}&=0,\\
		\vec{\mathbf{X}}^{n,s}&=0,
	\end{align}
\end{subequations}
where arrows indicate a discrete list of all $N^2$ mesh points or all $N_p$ particles as appropriate.
Together, (\ref{eq:disc_gov}) provides the prediction for the flow velocity, the particle velocities, and the particle positions at time step $n$ and Runge--Kutta sub-step $s$.
Appendix~\ref{app:discret_det} includes details on the discretization.

Because large-eddy simulation typically includes energy near the grid resolution, discretization errors are often comparable to the unclosed sub-grid contributions~\cite{CHOW2003,GHOSAL1996}.
Consequently, even a perfect model for the analytical forms of $\bm{\tau}^r$ and $\mathbf{u}^\prime$ may produce significant numerical error.
As formulated, the learned closures $\mathbf{h}_\mathrm{u}$ and $\mathbf{h}_\mathrm{p}$ should correct both sub-grid and discretization errors~\cite{sirignano_dpm_2020}.

To measure errors consistently with this discretization, we define discrete inner products that include the discretization parameters,
\begin{subequations}\label{eq:inner_product}
	\begin{align}
		\langle\vec{f},\vec{g}\,\rangle_u & \equiv  \sum_{n=1}^{N_t}\sum_{s=1}^4\sum_{i=1}^{2}{({\vec{f}_i}^{\; n,s})}^T{\vec{g}_i}^{\; n,s}\Delta x^2 \gamma_s\Delta t,\\
		\langle\vec{q},\vec{r}\,\rangle_p & \equiv  \sum_{n=1}^{N_t}\sum_{s=1}^4\sum_{i=1}^{2}{({\vec{q}_i}^{\; n,s})}^T{\vec{r}_i}^{\; n,s} \gamma_s\Delta t,
	\end{align} 
\end{subequations}
where weights $\gamma_s = [1/6,1/3,1/3,1/6]$ are consistent with the Runge--Kutta sub-steps~\cite{vishnampet_practical_2015}, the spacing of the (square) mesh is $\Delta x$, and the time step is $\Delta t$ for $\vec{f}, \vec{g} \in \mathbb{R}^{N^2\times N_t\times 4\times 2}$ and $\vec{q}, \vec{r} \in \mathbb{R}^{N_p\times N_t\times 4\times 2}$. 
Additionally, for discrete transforms $\widehat{f}, \widehat{g} \in \mathbb{C}^{N^2\times N_t\times 4}$ of the form $\widehat{f} = \sum_{\mathbf{x}} \vec{f}e^{-i\mathbf{x}\cdot\bm{\kappa}}$, which are used in some training schemes and analysis of results, we define 
\begin{gather}\label{eq:spectral_inner_product}
	\langle\widehat{f},\widehat{g}\rangle_\kappa \equiv \sum_{n=1}^{N_t}\sum_{s=1}^4\sum_{k=1}^{N} {\widehat{f}_k}^{*\; n,s}{\widehat{g}_k}^{\; n,s}\gamma_s\Delta\kappa^2\Delta t,
\end{gather} 
where $\widehat{f}^{\,*}$ is a complex conjugate. 

\section{Constrained optimization}\label{sec:PCO}
For the optimization problem (\ref{eq:apost_opt}), we use a equation-constrained approach~\cite{antil_brief_2018} with objective function $\mathcal{J}$, which is a generic $L_2$ mismatch between the neural-network-embedded large-eddy simulation predictions and trusted data for the same.
Various $\mathcal{J}$ are considered in section~\ref{sec:demonstration}.
The corresponding Lagrangian is
\begin{equation}\label{eq:lagrangian}
	\mathcal{L} = \mathcal{J} - \langle\vec{\mathbf{u}} ^{\dag}, \vec{\mathbf{N}}^{n,s} \rangle - \langle \vec{\mathbf{v}}_\mathrm{p}^{\dag}, \vec{\mathbf{M}}^{n,s}  \rangle- \langle \vec{\mathbf{x}}_\mathrm{p}^{ \dag}, \vec{\mathbf{X}}^{n,s}  \rangle,
\end{equation}
where $\vec{\mathbf{u}} ^{\dag}$, $\vec{\mathbf{v}}_\mathrm{p}^{\dag}$, and $\vec{\mathbf{x}}_\mathrm{p}^{\dag}$ are the Lagrange multipliers (adjoint variables) enforcing the governing equation.
The first-order optimality condition is
\begin{align}\label{eq:FOC}
	\nabla_{\vec{\mathbf{u}}}\mathcal{L} = \nabla_{\vec{\mathbf{v}}_\mathrm{p}}\mathcal{L} = \nabla_{\vec{\mathbf{x}}_\mathrm{p}}\mathcal{L} = \nabla_{\vec{\theta}_u}\mathcal{L} =	\nabla_{\vec{\theta}_p}\mathcal{L} = 0.
\end{align}
Gradients with respect to the Lagrange multipliers yield their adjoint governing equations. 
Expanding (\ref{eq:FOC}), we obtain
\begin{subequations}
	\begin{align}
		\vec{\mathbf{N}}^{\dagger n,s} \equiv\;& \langle\vec{\mathbf{u}} ^{\dag}, \nabla_{\vec{\mathbf{u}}}\vec{\mathbf{N}}^{n,s}  \rangle + \langle \vec{\mathbf{v}}_\mathrm{p}^{\dag}, \nabla_{\vec{\mathbf{u}}}\vec{\mathbf{M}}^{n,s}  \rangle -\nabla_{\vec{\mathbf{u}}}\mathcal{J}=0,\label{eq:u_adj_compact}\\
		\vec{\mathbf{M}}^{\dagger n,s} \equiv\;& \langle\vec{\mathbf{u}}^{\dag}, \nabla_{\vec{\mathbf{v}}_\mathrm{p}}\vec{\mathbf{N}}^{n,s}\rangle + \langle \vec{\mathbf{v}}_\mathrm{p}^{\dag}, \nabla_{\vec{\mathbf{v}}_\mathrm{p}}\vec{\mathbf{M}}^{n,s}  \rangle
		 +\langle \vec{\mathbf{x}}_\mathrm{p}^{ \dag}, \nabla_{\vec{\mathbf{v}}_\mathrm{p}}\vec{\mathbf{X}}^{n,s}\rangle - \nabla_{\vec{\mathbf{v}}_\mathrm{p}}\mathcal{J}=0\label{eq:vp_adj_compact},\\
		\vec{\mathbf{X}}^{\dagger n,s} \equiv\;&\langle\vec{\mathbf{u}} ^{\dag}, \nabla_{\vec{\mathbf{x}}_\mathrm{p}}\vec{\mathbf{N}}^{n,s}  \rangle + \langle \vec{\mathbf{v}}_\mathrm{p}^{\dag}, \nabla_{\vec{\mathbf{x}}_\mathrm{p}}\vec{\mathbf{M}}^{n,s}  \rangle +\langle \vec{\mathbf{x}}_\mathrm{p}^{ \dag}, \nabla_{\vec{\mathbf{x}}_\mathrm{p}}\vec{\mathbf{X}}^{n,s}\rangle - \nabla_{\vec{\mathbf{x}}_\mathrm{p}}\mathcal{J}=0\label{eq:xp_adj_compact},\\
		\nabla_{\vec{\mathbf{\theta}}_u}\mathcal{J} =\;&\langle \vec{\mathbf{u}} ^{\dag}, \nabla_{\vec{\mathbf{\theta}}_p}\vec{\mathbf{N}}^{n,s} \rangle\label{eq:grad_thetau},\\
		\nabla_{\vec{\mathbf{\theta}}_p}\mathcal{J} =\;& \langle\vec{\mathbf{v}}_\mathrm{p} ^{\dag}, \nabla_{\vec{\mathbf{\theta}}_p}\vec{\mathbf{M}}^{n,s}   \rangle .\label{eq:grad_thetap}
	\end{align}
\end{subequations} 
Equations (\ref{eq:u_adj_compact}), (\ref{eq:vp_adj_compact}), and (\ref{eq:xp_adj_compact}) are the adjoint system, which is solved numerically, while (\ref{eq:grad_thetau}) and (\ref{eq:grad_thetap}) are the gradients used to optimize weights, as in (\ref{eq:apost_opt}).   
Using the discrete governing equations as constraints leads to discrete-exact adjoints that provide machine-precision gradients of the training parameters~\cite{vishnampet_practical_2015}.
The full derivation included in appendix~\ref{app: adj_der} provides final time conditions consistent with integrating the adjoint system in reverse time~\cite{antil_brief_2018}.

\section{Training data}\label{sec:training_data}
\begin{figure}
	{\centering  
	\includegraphics[width=0.8\linewidth]{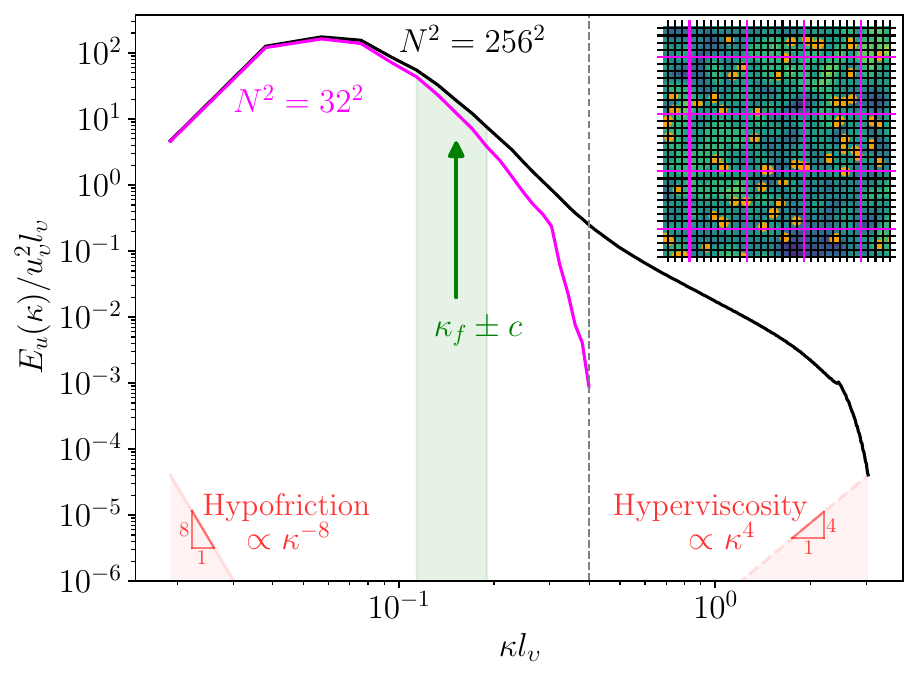}\par}
	\captionsetup{justification=justified,singlelinecheck=false}
	\caption{Turbulence kinetic energy before and after filtering, with indicated hypofriction, hyperviscosity, and forcing wavenumber ranges (see section~\ref{sec:training_data}). It is normalized by the dissipation length scale $l_\upsilon = Z_\omega^{1/6}/\upsilon^{1/2}$ and dissipation velocity scale $u_\upsilon = \upsilon^{1/2} Z_\omega^{1/6}$, where enstrophy dissipation rate is $Z_\omega = \upsilon\langle|\nabla\vec{\bm{\omega}}|^2\rangle_x$~\cite{bracco_reynolds-number_2010}. An inset showing a subregion of side length $L=0.125$ visualizes the mesh spacings used.}\label{fig:dns_spec}
\end{figure}

Nothing about the formulation requires DNS data, yet it provides an attractively precise option for developing the approach.
(In section~\ref{sec:noise} we also introduce noise-like errors to model less precise data.)
An isotropic velocity field was initialized using standard procedures~\cite{rogallo_numerical_1981} in a periodic square domain of length $L=50$ and with the time step $\Delta t_\mathrm{DNS}=3.54\times10^{-3}$ on a uniform $N_\mathrm{DNS}^2 = 256^2$ mesh.
Hyperviscosity $\upsilon  = 5.50\times 10^2$ and hypofriction $\alpha = 2.82\times 10^{-10}$ values were chosen to obtain the spectra in figure~\ref{fig:dns_spec}.
Once the variance of the flow velocity became stationary, within 1\% of its mean over $2000\tau_\omega$, where $\tau_\omega =2/\omega_\mathrm{rms}$ is the enstrophy (turnover) time scale based on the root-mean square vorticity $\omega_\mathrm{rms}$, then $N_p = 2000$ particles were randomly distributed in the domain.
The particle relaxation time $\tau_p$ was set for Stokes number $St = \tau_p/\tau_\eta=0.8$, where $\tau_\eta = 1/\sqrt{2S}$ is the strain-rate time scale with $S$ the mean trace of the strain-rate tensor, which is known to approximately maximize particle clustering in two-dimensional turbulence~\cite{boffetta_large_2004,goto_self-similar_2006}.
The coupled system was advanced until both fluid and particle statistics satisfied the same $1\%$ stationarity criterion used for the initial flow development.
Then, data were saved for $1000\tau_\omega$.
Five different initial conditions from different random seeds were used to provide more training data.
The Reynolds number was $Re = u_\mathrm{rms} L^3/\upsilon = 3.7\times 10^3$ for all cases~\cite{lamorgese_direct_2004}. 

For training, the DNS flow velocity, $\vec{\mathbf{U}}$, was filtered using (\ref{eq:filter}) with the box filter of width $\Delta = \Delta x_\mathrm{LES} = 8\Delta x_\mathrm{DNS}$, which amounts to a local average:
\begin{equation}\label{eq:kernel}
	G(\mathbf{x}^\prime,\mathbf{x}) = \begin{cases}
		\frac{1}{\Delta^2} & \text{if} \; |\mathbf{x}-\mathbf{x}^\prime|\leq \frac{\Delta}{2}, \\
		0 & \text{otherwise.}
	\end{cases}x
\end{equation}
After filtering, $\overline{\mathbf{U}}$ was downsampled to the coarse $N_\text{LES}$ grid and projected onto a divergence-free manifold~\cite{sirignano_dpm_2020}: $\vec{\mathbf{W}} = \mathbf{P} \overline{\mathbf{U}}$, where $\mathbf{P}$ is the projection operator that guarantees a discrete solenoidal field: $\mathbf{D}\vec{\mathbf{W}} = 0$ (see appendix~\ref{app:projection}). 

Figure~\ref{fig:dns_spec} compares the spectra before and after this pre-processing. 
Both hypofriction and hyperviscosity act in relatively local wavenumber ranges, leaving the forced scales essentially unchanged.  
The forcing wavenumbers are within the represented scales, and the processed spectrum matches the DNS peak energy.

\section{Neural network closure}\label{sec:NN_design}
\subsection{Momentum conservation}\label{sec:mom_cons_NN}
Embedding $\vec{\mathbf{h}}_\mathrm{u}$ in divergence form in (\ref{eq:cont_NN_NS}) ensures that it adds no momentum, which is also preserved by our discretization.
In the absence of $\vec{\mathbf{h}}_\mathrm{p}$ in (\ref{eq:cont_NN_MR}), the form of the  operator $\mathbf{B}^T$ guarantees conservation of momentum. 
However, to exactly conserve momentum with $\vec{\mathbf{h}}_\mathrm{p}$ added, its sum over all particles must be zero.
This is enforced by subtracting its mean:
\begin{equation}
	\sum^{N}_{i=1} \vec{\mathbf{F}}_{\mathrm{p}_i} = \sum_{j=1}^{N_p} \Big[m_p\frac{d \vec{\mathbf{v}}_{\mathrm{p}_j}}{dt} + \vec{\mathbf{h}}_{\mathrm{p}_j}\Big]-\frac{1}{N_p}\sum_{j=1}^{N_p}\vec{\mathbf{h}}_{\mathrm{p}_j}.
\end{equation}
This minute correction each time step is equivalent to an orthogonal projection of $\vec{\mathbf{h}}_{\mathrm{p}}$ onto the zero-sum subspace (appendix~\ref{app:hp_projection}), analogous to center-of-mass drift removal when applying a thermostat in molecular dynamics simulations~\cite{allen_computer_2017,harvey_flying_1998}.

\subsection{Neural-network inputs}
We take $\vec{\mathbf{h}}_\mathrm{u}$ to depend on the velocity differences between the target meshpoint and its eight nearest neighbors, so its input across all mesh points is $\vec{\bm{\phi}}_{u}\in\mathbb{R}^{8\times N^2\times 2}$.
The implementation constructs this via
\begin{equation}
	\vec{\bm{\phi}}_{u_k} = \vec{\mathbf{u}}-{\big(\mathbf{H}_\mathrm{u}\vec{\mathbf{u}}\big)}_k \quad \mathrm{for\: neighbors} \quad k=1,\ldots, 8,
\end{equation}
at each time step, where $\mathbf{H}_\mathrm{u}\in\mathbb{R}^{8\times N^2\times N^2}$ is an operator that selects the nearest eight neighbor points.
Using velocity differences ensures Galilean invariance.
Tensorial symmetry is imposed on $\vec{\mathbf{h}}_\mathrm{u}$, but other symmetries such as rotational invariance are not enforced, though they could be.
The $\vec{\mathbf{h}}_\mathrm{p}$ inputs in (\ref{eq:cont_NN_MR}) include the velocity differences 
\begin{equation}
	\bm{\Delta} \vec{\mathbf{u}} = \begin{bmatrix}
		\vec{\mathbf{u}}_{i+1,j}-\vec{\mathbf{u}}_{i,j}\\
		\vec{\mathbf{u}}_{i,j+1}-\vec{\mathbf{u}}_{i,j}
	\end{bmatrix},
\end{equation}
at the faces of the mesh cell it occupies. This is represented as $\mathbf{H}_\mathrm{p}\bm{\Delta}\vec{\mathbf{u}}$, where the $\mathbf{H}_\mathrm{p}\in\mathbb{R}^{4\times N_p\times N^2}$ operator selects these neighbors. 
Additionally, it includes the particle lag velocities relative to the mesh cell faces, $\vec{\mathbf{v}}_\mathrm{p}-{\big(\mathbf{H}_\mathrm{p}\vec{\mathbf{u}}\big)}_k$, and its location relative to them, $[\vec{\mathbf{x}}_\mathrm{p}-{(\mathbf{H}_\mathrm{p}\vec{\mathbf{x}})}_k]/\Delta \mathbf{x}$: 
\begin{equation}
	\vec{\bm{\phi}}_{p_k} = {\bigg[{(\mathbf{H}_\mathrm{p}\bm{\Delta}\vec{\mathbf{u}})}_k,\vec{\mathbf{v}}_\mathrm{p}-{(\mathbf{H}_\mathrm{p}\vec{\mathbf{u}})}_k, \frac{\vec{\mathbf{x}}_\mathrm{p}-{(\mathbf{H}_\mathrm{p}\vec{\mathbf{x}})}_k}{\Delta \mathbf{x}}\bigg]}^T \quad\mathrm{for\: mesh\: points}\quad k=1,\ldots, 4,
\end{equation}
where $\vec{\bm{\phi}}_{p}\in\mathbb{R}^{16\times N_p\times 2}$.
Using the relative location as input should allow the network to learn proximity-weighted corrections for particle dynamics that counter discretization errors synergetically with the $\mathcal{J}$ objective. 

\subsection{Neural-network architecture}\label{sec:NN model}
We use identical established feedforward deep neural networks for $\vec{\mathbf{h}}_\mathrm{u}(\vec{\bm{\phi}}_u;\vec{\theta}_{u})$ and $\vec{\mathbf{h}}_\mathrm{p}(\vec{\bm{\phi}}_p;\vec{\theta}_{p})$.
For the $i^\mathrm{th}$ component of input $\vec{\phi}_{u_i}$~\cite{sirignano_dpm_2020}:
\begin{subequations}\label{eq:NN_arch}
	\begin{align}
		H^1 &= \sigma(W^1\vec{\phi}_{u_i}+b^1),\\
		H^2 &= \sigma(W^2H^1+b^2),\\
		H^3 &= G^1 \odot H^2, \qquad \qquad \text{where} \qquad G^1 = \sigma(W^5\vec{\phi}_{u_i} + b^5), \\
		H^4 &= \sigma(W^3H^3+b^3),\\
		H^5 &= G^2 \odot H^4, \qquad \qquad \text{where} \qquad G^2 = \sigma(W^6\vec{\phi}_{u_i} + b^6), \\
		\vec{h}_{\mathrm{u}_i}(\vec{\phi}_{u_i};\vec{\theta}_{u}) &= W^4H^5+b^4,
	\end{align}
\end{subequations}
where $\sigma(\cdot) = \tanh(\cdot)$, $\odot$ is element-wise multiplication, and the parameters to be optimized are
\begin{equation*}
\vec{\theta}_{u} = \{W^1,W^2,W^3,W^4,W^5,W^6,b^1,b^2,b^3,b^4,b^5,b^6\}.
\end{equation*} 
Each layer consists of $20$ hidden units, with a uniform Xavier~\cite{glorot_understanding_2010} initialization.
Additional hidden units provided negligible improvements. 
This architecture has been shown effective for single-phase turbulence~\cite{macart_embedded_2021,sirignano_dpm_2020}.
The gates $\mathbf{G}^1$ and $\mathbf{G}^2$ serve as input-dependent feature selectors for parameters.
Such gates are expected to improve generalization as they allow the neural-network to tune the weights to different regimes.
The same architecture is used for $\vec{\mathbf{h}}_\mathrm{p}(\vec{\bm{\phi}}_p;\vec{\theta}_{p})$.

\subsection{Training}\label{sec:training}
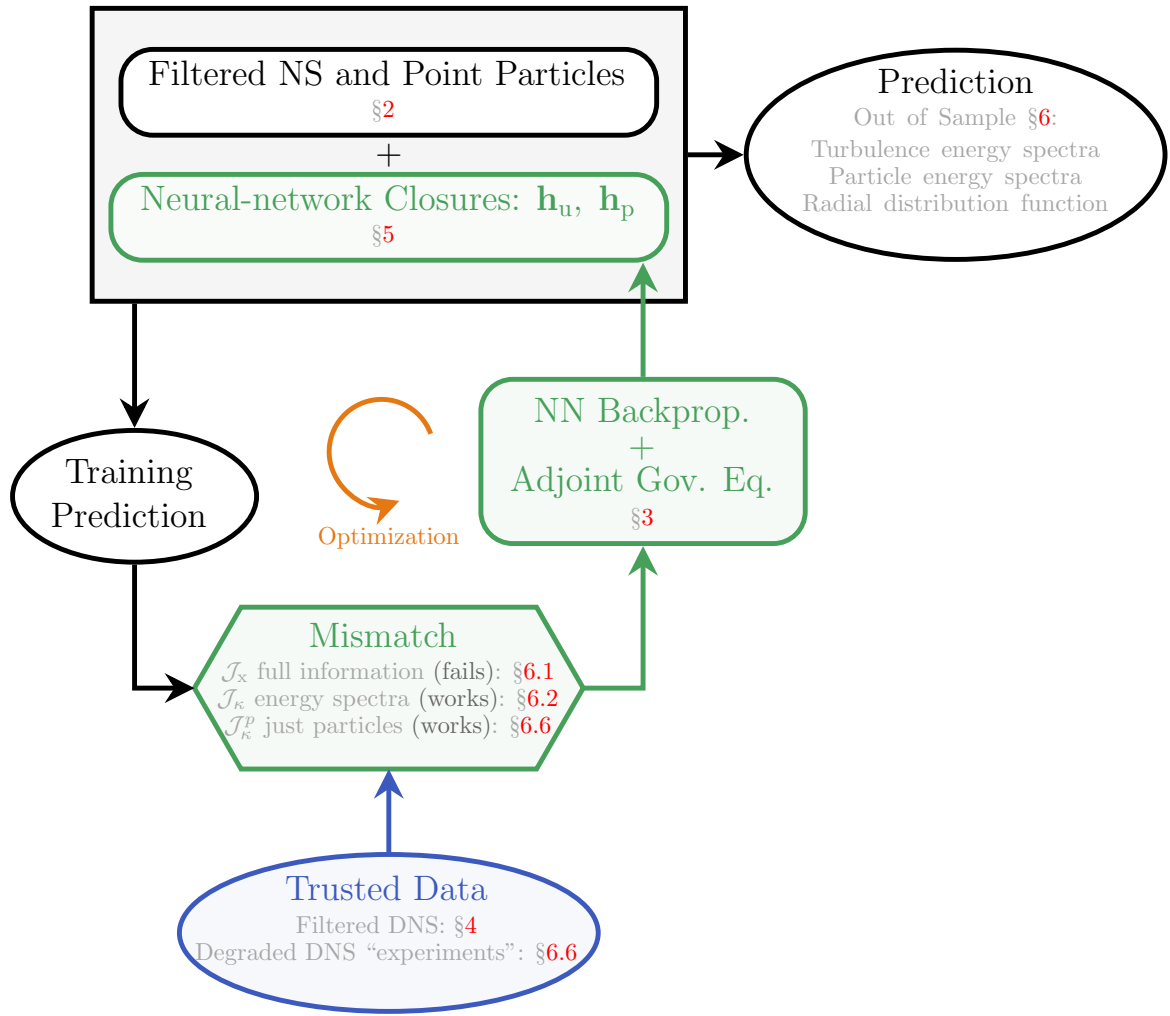
\begin{figure}
  \centering
  \begingroup\setstretch{1}%
    \renewcommand{\Large}{\fontsize{14.4}{18}\selectfont}%
    \renewcommand{\normalsize}{\fontsize{10}{11}\selectfont}%
    \resizebox{0.95\linewidth}{!}{\begin{tikzpicture}[font=\rmfamily\Large,
  bigbox/.append style={line width=2pt},
  rrect/.append style={line width=2pt},
  oval/.append style={line width=2pt},
  polyG/.append style={line width=2pt}
]

\node[bigbox, fill=black!4] (dpm) at (8.4,2.5) {};

\coordinate (plusPos) at ($(dpm.center)+(0,0)$);
\node[font=\Large] (plus) at (plusPos) {$+$};

\def\plusPad{7pt}

\node[rrect, draw=black,fill=white, text=black,
      minimum width=5.7cm, anchor=south] (pde)
  at ($(plusPos)+(0,\plusPad)$) {%
  \begin{tabular}{c}
    {\:Filtered NS and Point Particles}\\[-5pt]
    {\normalsize\color{mygray}\S\ref{sec:formulation}}
  \end{tabular}
};

\node[rrect, anchor=north,
      minimum width=5.7cm, fill=mygreen!4, text opacity=1] (closure)
  at ($(plusPos)+(0,-\plusPad)$) {%
  \begin{tabular}{c}
    {\textcolor{mygreen}{\;Neural-network Closures: $\mathbf{h}_\mathrm{u}, \;\mathbf{h}_\mathrm{p}$}}\\[-5pt]
    {\normalsize\color{mygray}\S\ref{sec:NN_design}}
  \end{tabular}
};

\coordinate (trainOut) at ($(dpm.south west)!0.075!(dpm.south east)$);
\coordinate (greenOut) at ($(dpm.south west)!0.925!(dpm.south east)$);
\coordinate (xalign)   at (trainOut |- closure.center);
\coordinate (xalignR)  at (greenOut |- closure.center);
\coordinate (closureIn) at ($(closure.south west)!(greenOut)!(closure.south east)$);

\node[oval, anchor=west,
      minimum width=6cm, minimum height=3cm,
      inner sep=0pt] (pred)
  at ($(dpm.east)+(0.8cm,0)$) {};

\node[anchor=center, align=center, text width=6.2cm] 
  at ([xshift=0mm,yshift=2mm]pred.center) {%
  \begin{tabular}{c}
    Prediction\\[-5pt]
    {\normalsize\color{mygray}Out of Sample \S\ref{sec:demonstration}:}\\
    {\normalsize\color{mygray}\begin{tabular}{@{}c@{}}
      Turbulence energy spectra\\
      Particle energy spectra\\
      Radial distribution function
    \end{tabular}}
  \end{tabular}
};

\node[oval] (train) at ($(xalign)+(0,-4)$) {%
  \begin{tabular}{@{}c@{}}
    Training\\
    Prediction
  \end{tabular}
};

\coordinate (misY)   at ($(xalign)+(0,-6.75)$);     
\coordinate (misPos) at (dpm.center |- misY);    

\node[polyG, fill=mygreen, fill opacity=0.05, text opacity=1] (mis) at (misPos) {};

\node[align=center] at ([yshift=4pt]mis.center) {%
  \begin{tabular}{@{}c@{}}
    {\color{mygreen}Mismatch}\\
    \;{\normalsize\color{mygray}\begin{tabular}{@{}c@{}}
      $\mathcal{J}_{\mathrm{x}}$ full information \textcolor{mydarkgray}{(fails)}: \S\ref{sec:full_obj}\\
      $\mathcal{J}_{\kappa}$ energy spectra \textcolor{mydarkgray}{(works)}: \S\ref{sec:spec_obj}\\
      $\mathcal{J}_{\kappa}^{p}$ just particles \textcolor{mydarkgray}{(works)}: \S\ref{sec:robustness}
    \end{tabular}}
  \end{tabular}
};

\node[rrect, draw=mygreen, text=mygreen, minimum width=2cm, minimum height=1.6cm,
      fill=mygreen, fill opacity=0.05, text opacity=1]
  (adjbp) at ($(xalignR)+(0,-3.5)$) {%
  \;\begin{tabular}{c}
    \\[-15pt]
    NN Backprop.\\[-4pt]
    {\Large $+$}\\[-4pt]
    Adjoint Gov. Eq.\\[-4pt]
    {\normalsize\color{mygray}\S\ref{sec:PCO}}
  \end{tabular}
};

\node[oval, draw=myblue, fill=myblue!4,
      minimum width=6cm, minimum height=2.25cm,
      inner sep=0pt] (data)
  at ($(mis.south)+(0,-2.35)$) {};

\node[align=center, text=myblue]
  at ([xshift=0.5mm,yshift=2mm]data.center) {%
  \begin{tabular}{c}
    Trusted Data\\
    {\normalsize\color{mygray}\begin{tabular}{@{}c@{}}
      Filtered DNS: \S\ref{sec:training_data}\\
      \begin{tabular}{@{}c@{}l@{}}
        Degraded DNS ``experiments" & {: \S\ref{sec:robustness}}
      \end{tabular}
    \end{tabular}}
  \end{tabular}
};

\draw[bigarrow, color=myblue] (data.north) -- (mis.south);

\draw[bigarrow] (trainOut) -- (train.north);

\draw[bigarrow] (train.south) |- (mis.west);

\coordinate (misAdjElbow) at (mis.east -| adjbp.center);

\draw[bigarrow, color=mygreen]
  (mis.east) -- (misAdjElbow) -- (adjbp.south);

\draw[bigarrow, color=mygreen]
  (adjbp.north) -- (greenOut) -- (closureIn);

\draw[bigarrow] (dpm.east) -- (pred.west);

\coordinate (stoLabel) at (dpm.center |- { (0,-3) });
\node[font=\normalsize\color{myorange}, align=center] at (stoLabel)
  {Optimization};

\coordinate (stoArcC) at ($(dpm.center)+(0.6cm,-4cm)$);
\draw[bigarrow, color=myorange, shift={(1cm,0)}]
  (stoArcC) arc[start angle=20, end angle=290, radius=0.75cm];

\end{tikzpicture}}%
  \endgroup
  \captionsetup{justification=justified,singlelinecheck=false}
  \caption{Workflow diagram showing the model components and learning loop for the adjoint-based learning of neural-network closures.}\label{fig:flowchart}
\end{figure}

A schematic of the training algorithm is shown in figure~\ref{fig:flowchart}.
The ADAM optimizer~\cite{kingma_adam_2014} uses stochastic gradient descent with minibatches of size $N_b=100$. 
Each mini-batch constituent has an initial condition randomly sampled from $1.2\times 10^5$ available realizations from the processed DNS data in section~\ref{sec:training_data}. 
The training proceeds as follows:
\begin{enumerate}
\item Initialize $\vec{\theta}_{u}^0$ and $\vec{\theta}_{p}^0$. The last layer is set to zero to guarantee $\vec{\mathbf{h}}_\mathrm{u} = 0$ and $\vec{\mathbf{h}}_\mathrm{p} = 0$ at the start of optimization to prevent extreme model outputs. 
\item For each training epoch $z = 1, 2, \ldots, N_e$:
\begin{itemize}
	\item Randomly select $N_b$ initial conditions at time $t^b$ from the post-processed DNS data in section~\ref{sec:training_data};
	\item Time-integrate the discrete governing equations (\ref{eq:disc_gov}) for $t\in[t^b,t^b+N_t\Delta t_\mathrm{LES}]$, for $b=1,\ldots,N_b$ and $\Delta t_\mathrm{LES} = 8\Delta t_\mathrm{DNS}$;
	\item Compute $\mathcal{J}^{z, b}$;
	\item Solve the adjoint equations (\ref{eq:u_adj_compact}), (\ref{eq:vp_adj_compact}), and (\ref{eq:xp_adj_compact}) for $t=[t^b+N_t\Delta t_\mathrm{LES}, t^b]$;
	\item Form the gradients (\ref{eq:grad_thetau}) and (\ref{eq:grad_thetap})
	\begin{subequations}
			\begin{align}
			& \nabla_{ \vec{\theta}_{u}^z}\mathcal{J}^{z,b}= \Big\langle \vec{\mathbf{u}}^\dagger, \mathbf{P}  \nabla_{\vec{\theta}_{u}^z}(\mathbf{D}\vec{\mathbf{h}}_\mathrm{u})\Big\rangle, \\ 
			& \nabla_{ \vec{\theta}_{p}^z}\mathcal{J}^{z,b}= \Big\langle\vec{\mathbf{v}}_\mathrm{p}^{\dagger},  \nabla_{ \vec{\theta}_{p}^z}\vec{\mathbf{h}}_\mathrm{p} \Big\rangle;
		\end{align}
	\end{subequations}
	\item Average over the $N_b$ minibatches
	\begin{subequations}
			\begin{align}
			\nabla_{ \vec{\theta}_{u}^z}\mathcal{J}^z &= \frac{1}{N_b}\sum_{b=1}^{N_b}\nabla_{ \vec{\theta}_{u}^z}\mathcal{J}^{z,b},\\
			\nabla_{ \vec{\theta}_{p}^z}\mathcal{J}^z&=\frac{1}{N_b}\sum_{b=1}^{N_b}\nabla_{ \vec{\theta}_{p}^z}\mathcal{J}^{z,b}; 
		\end{align}
	\end{subequations}
	\item And, update parameters using ADAM~\cite{kingma_adam_2014}
	\begin{subequations}\label{eq:ADAM}
		\begin{align}
			\vec{m}^z  &= \beta_1 \vec{m}^{z-1} +(1-\beta_1)\nabla_{\vec{\theta}_{u}^z}\mathcal{J}^z, \\
			\vec{v}^{\,z} &= \beta_2 \vec{v}^{\,z-1}+(1-\beta_2)\Big(\nabla_{\vec{\theta}_{u}^z}\mathcal{J}^z\odot\nabla_{\vec{\theta}_{u}^z}\mathcal{J}^z\Big),\\
			\widehat{m}^z  &= \frac{\vec{m}^z}{1-{(\beta_1)}^z},\\ 
			\widehat{v}^{\,z}  &= \frac{\vec{v}^{\,z}}{1-{(\beta_2)}^z},\\ 
			\vec{\theta}_{u}^z  &= \vec{\theta}^{z-1}_u - \frac{c_\mathrm{lr} }{\sqrt{\widehat{v}^{\,z}}+\varepsilon}\odot\widehat{m}^z.
		\end{align}
	\end{subequations}
\end{itemize} 
\end{enumerate}
The ADAM hyperparameters are $\beta_1 = 0.9$, $\beta_2 = 0.999$, $\varepsilon = 10^{-8}$, $\vec{m}^0 = \vec{v}^{\,0} = 0$, and the learning rate is $c_\mathrm{lr} = 0.001$. 
The particle closure is trained correspondingly. 

Choosing how many time steps $N_t$ to train over depends on the regime.
The chaotic character of particle-laden turbulence makes the solution sensitive to perturbations, so $\nabla_{\vec{\mathbf{\theta}}_u}\mathcal{J}$ and $\nabla_{\vec{\mathbf{\theta}}_p}\mathcal{J}$ grow rapidly with increasing integration (reverse) time~\cite{CHUNG2022111077}. 
Integration time step numbers $N_t=10$, $20$, $40$, $80$, and $160$ were tested to determine training time horizons.
Only marginal improvements were observed beyond $N_t=40$. 
For $N_t =160$, gradients became too large for convergence with these parameters.
Thus, $N_t=40$ was selected for our cases.
This is further discussed in section~\ref{sec:conc}.

\section{Training and prediction results}\label{sec:demonstration}
\subsection{Full state mismatch objective $\mathcal{J}_\mathrm{x}$}\label{sec:full_obj}
When we have a trusted solution, the obvious training target is simply the time-dependent fields:
\begin{align}\label{eq:pointwise_loss}
	\mathcal{J}_\text{x} =  C_u\langle \Delta\vec{\mathbf{u}}, \Delta\vec{\mathbf{u}} \rangle_u + C_{v_p}\langle \Delta\vec{\mathbf{v}}_\mathrm{p}, \Delta\vec{\mathbf{v}}_\mathrm{p} \rangle_p + C_{x_p}\langle \Delta\vec{\mathbf{x}}_\mathrm{p}, \Delta\vec{\mathbf{x}}_\mathrm{p}\rangle_p,
\end{align}
where $\Delta\vec{\mathbf{u}}  = \vec{\mathbf{u}} - \vec{\mathbf{W}}$, $\Delta\vec{\mathbf{v}}_\mathrm{p} = \vec{\mathbf{v}}_\mathrm{p} - \vec{\mathbf{V}}_\mathrm{p}$, $\Delta\vec{\mathbf{x}}_\mathrm{p} = \vec{\mathbf{x}}_\mathrm{p} - \vec{\mathbf{X}}_\mathrm{p}$ and $\vec{\mathbf{W}}$ is the pre-processed DNS flow of section~\ref{sec:training_data}.
DNS particle data $\vec{\mathbf{V}}_\mathrm{p}$, and $\vec{\mathbf{X}}_\mathrm{p}$ do not require processing.
Coefficients $C_u$, $C_{v_p}$, and $C_{x_p}$ make the contributions of each term the same for the initial condition used.

\begin{figure} 
 	\centering
 	\includegraphics[width=0.7\linewidth]{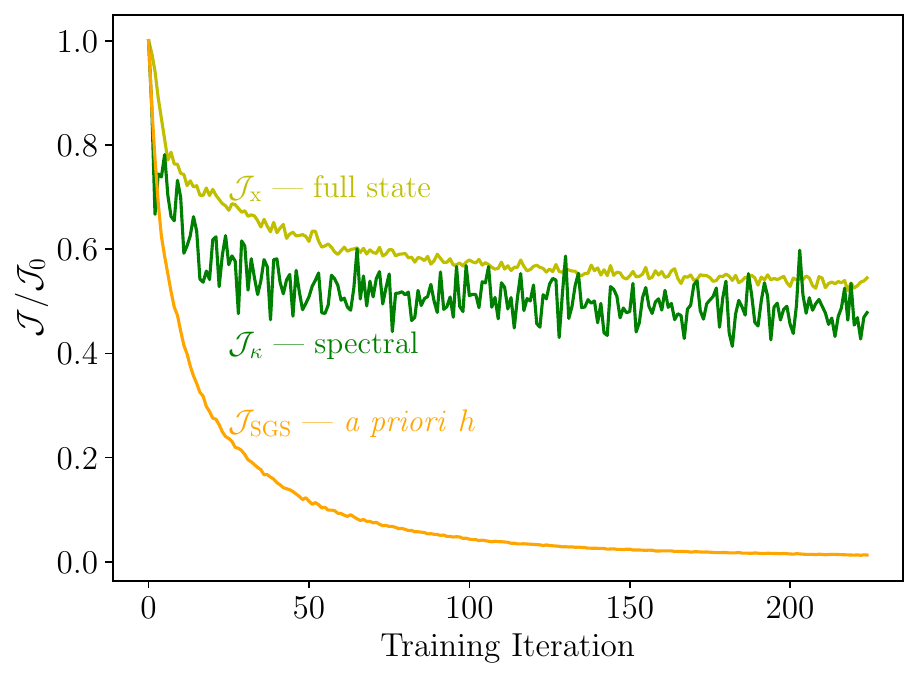}
	\captionsetup{justification=justified,singlelinecheck=false}
 	\caption{Training for full state $\mathcal{J}_\mathrm{x}$ (\ref{eq:pointwise_loss}), spectral $
	\mathcal{J}_\kappa$ (\ref{eq:spectral_loss}), and the \textit{a priori} closure match $\mathcal{J}_\mathrm{SGS}$ (\ref{eq:apiori_loss}).}\label{fig:training_curve}
\end{figure}

Note that we consider this $\mathcal{J}=\mathcal{J}_\mathrm{x}$ first because it is the most direct, although we will see subsequently that other options will significantly outperform it.
Once converged as in figure~\ref{fig:training_curve}, predictions are then made for out-of-sample initial conditions generated per section~\ref{sec:training_data}.
The predicted turbulence kinetic energy spectrum,
\begin{equation}\label{eq:Ek}
	\vec{E}_u(\vec{\kappa}) = \sum_{\vec{\bm{\kappa}}\in\mathcal{S}_l}\frac{1}{2N_l}|\widehat{\mathbf{u}}(\vec{\bm{\kappa}})|^2,
\end{equation}
is compared to the no-model LES and the pre-processed DNS results, where $\widehat{\mathbf{u}}$ is the discrete Fourier transform (DFT) of $\vec{\mathbf{u}}$, $N_l$ is the number of discrete wavenumbers within $\Delta\kappa$ shell $l$, and $\vec{\bm{\kappa}} = (2\pi i/L,2\pi j/L)$ with $i,j = -N/2,\ldots, N/2-1$ ($N$ even).
The radial bins used in computing the spectrum are $\mathcal{S}_l = \{\vec{\bm{\kappa}}: (l-1/2)\Delta\kappa\leq|\vec{\bm{\kappa}}|<(l+1/2)\Delta\kappa\}$ with $\Delta\kappa = 2\pi/L$ and $l = 0,\ldots, N/2L$.

\begin{figure}
	\begin{subfigure}{0.8\textwidth}
		\includegraphics[width=\linewidth]{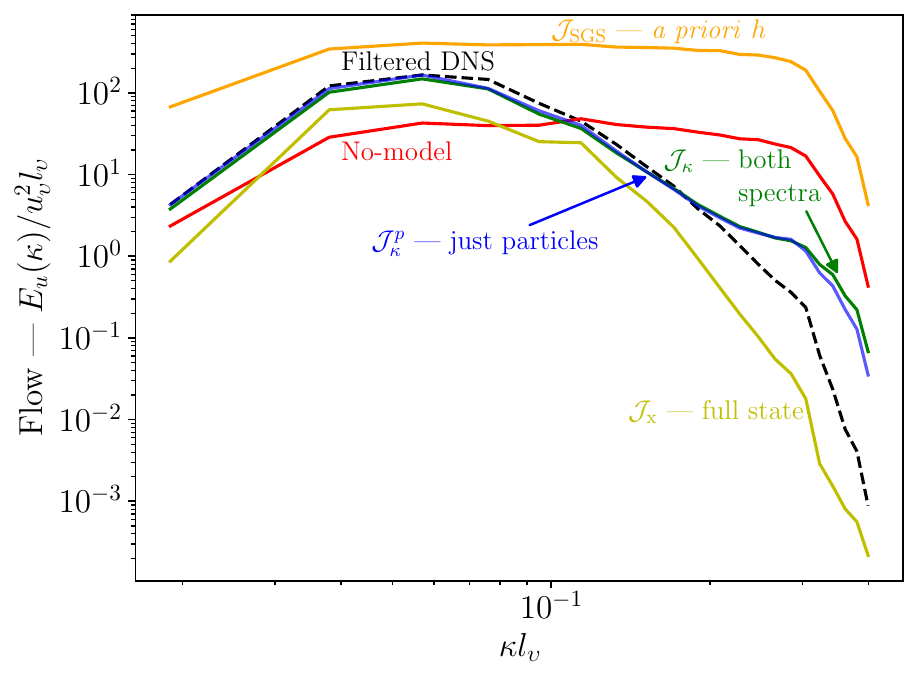}
		\caption{Turbulence kinetic energy}\label{fig:tke_point}
	\end{subfigure}
	\begin{subfigure}{0.8\textwidth}
		\includegraphics[width=\linewidth]{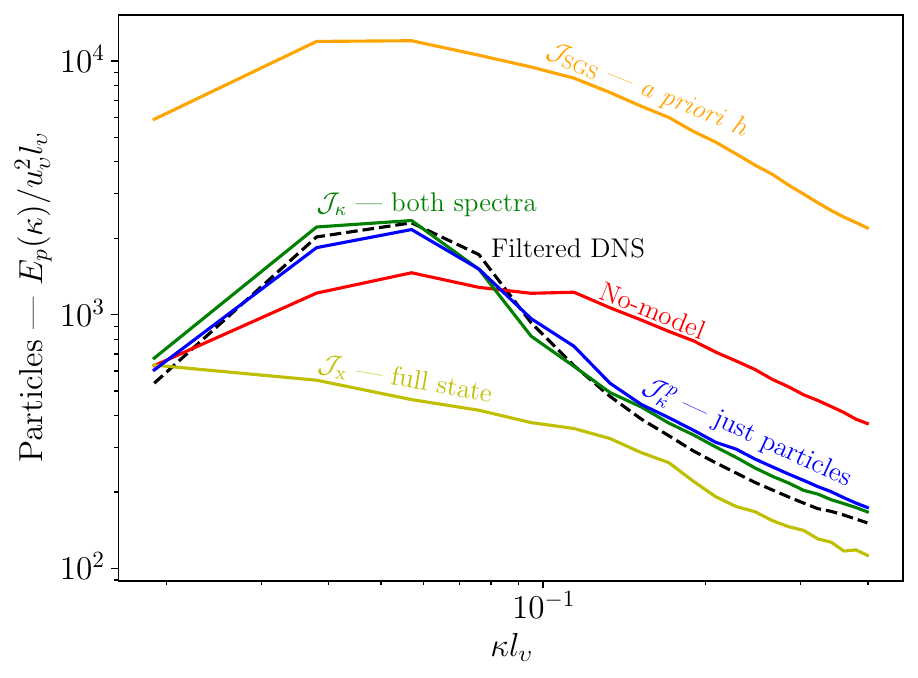}
		\caption{Particle kinetic energy}\label{fig:vpke_point}
	\end{subfigure}
	\captionsetup{justification=justified,singlelinecheck=false}
	\caption{Time-averaged spectra for the full state $\mathcal{J}_\mathrm{x}$ (\ref{eq:pointwise_loss}), both spectra $\mathcal{J}_\kappa$ (\ref{eq:spectral_loss}), \textit{a priori} $\mathcal{J}_\mathrm{SGS}$ (\ref{eq:apiori_loss}), and particle spectra $\mathcal{J}_\kappa$ (\ref{eq:particle_loss}) loss functions.}\label{fig:pointwise}
\end{figure}

Figure~\ref{fig:pointwise}(a) shows spectra (\ref{eq:Ek}) averaged over the statistically stationary period $t/t_{\omega}\in[250,1000]$.
Missing significant dissipation, the no-model LES reaches an equilibrium with a qualitatively incorrect, nearly uniform energy spectrum. 
(Note that a dissipative numerical method would show the opposite behavior, with under-resolution causing excessive dissipation and a rapidly decaying spectrum.)
The mean turbulence kinetic energy $k_u=\frac{1}{2}\langle \mathbf{u}\cdot\mathbf{u} \rangle_x$ is more than twice the filtered DNS value (table~\ref{tab:mean_QoI}).
In contrast, the $\mathcal{J}_{\text{x}}$ neural-network LES has qualitatively reasonable energy distribution but is overly dissipative, yielding only half the filtered DNS turbulence kinetic energy.

The irregular particle positions warrant analysis with a non-uniform discrete Fourier transform (NUDFT):
\begin{equation} \label{eq:NUFT}
	\widehat{\mathbf{v}}_\mathrm{p}(\vec{\bm{\kappa}}) = \frac{1}{N_p}\sum_{j=1}^{N_p}\vec{\mathbf{v}}_{\mathrm{p}_j}e^{-i\vec{\mathbf{x}}_{\mathrm{p}_j}\cdot\vec{\bm{\kappa}}}.
\end{equation}
The same discrete wavenumbers $\vec{\bm{\kappa}}$ supported by the LES mesh were used.
Similar transforms are commonly used in grid-to-particle interpolation schemes for particle-laden flows~\cite{beylkin_fast_1995,carbone_application_2018,tom_how_2022,van_hinsberg_efficiency_2012}.
Using (\ref{eq:NUFT}), the particle kinetic energy spectrum is 
\begin{equation}\label{eq:ep}
	\vec{E}_p^\circ(\vec{\kappa}) = \sum_{\vec{\bm{\kappa}}\in\mathcal{S}_l}\frac{1}{2N_l}|\widehat{\mathbf{v}}_\mathrm{p}(\vec{\bm{\kappa}})|^2.
\end{equation} 
However, since discrete orthogonality depends on uniform mesh spacing, the NUDFT does not satisfy Parseval's theorem, so the nominal energy (\ref{eq:ep}) is simply rescaled to match the known mean particle kinetic energy $k_p = \frac{1}{2}\langle \mathbf{v}_\mathrm{p}\cdot\mathbf{v}_\mathrm{p} \rangle_p$ for the same case:
\begin{equation}\label{eq:ep_rescaled}
	\vec{E}_p(\vec{\kappa}) = \frac{k_p}{\sum_{\vec{\kappa}}\vec{E}_p^\circ(\vec{\kappa})} \vec{E}_p^\circ(\vec{\kappa}).
\end{equation} 

Spectra in figure~\ref{fig:pointwise}(b) show that, as expected, the filtered DNS particle kinetic energy peaks near the most energetic flow scales. 
The no-model LES has an excess of large-wavenumber particle energy, consistent with the flow spectrum $\vec{E}_u(\vec{\kappa})$, yielding a mean kinetic energy $k_p$ that is about 40\% higher than DNS.\@ 
The $\mathcal{J}_\mathrm{x}$ neural-network-LES predicts roughly half of the DNS mean particle kinetic energy with a distorted shape.

\begin{table}
	\centering
	\captionsetup{justification=justified,singlelinecheck=false}
 	\begin{tabular}{lccc}\hline
 		LES Case		& $k_u/k_{u_\mathrm{DNS}}$ & $k_p/k_{p_\mathrm{DNS}}$ \\ \hline\hline
 		No-model    															& 2.37    & 1.40    \\
 		Full state $\mathcal{J}_\mathrm{x}$ (\S\ref{sec:full_obj})    			& 0.39    & 0.47    \\ 
		Both spectra $\mathcal{J}_\kappa$  	(\S\ref{sec:spec_obj})				& 0.92    & 1.04    \\ 
		\textit{A priori} $\mathcal{J}_\mathrm{SGS}$ (\S\ref{sec:apriori})	 	& 24.70   & 10.10   \\
		Particle spectra $\mathcal{J}_\kappa^p$  (\S\ref{sec:part_obj})   		& 0.95    & 1.03    \\ 
		One component $\mathcal{J}_\kappa^{p_1}$  (\S\ref{sec:one_vp_comp}) 	& 0.95    & 1.00   \\\hline     
 	\end{tabular}
	\caption{Time averaged ($t\in[250,1000]t_{\omega}$) turbulence kinetic energy and particle kinetic energy normalized by the trusted DNS data.}\label{tab:mean_QoI}
\end{table}
\begin{figure}
		\centering
		\includegraphics[width=0.8\linewidth]{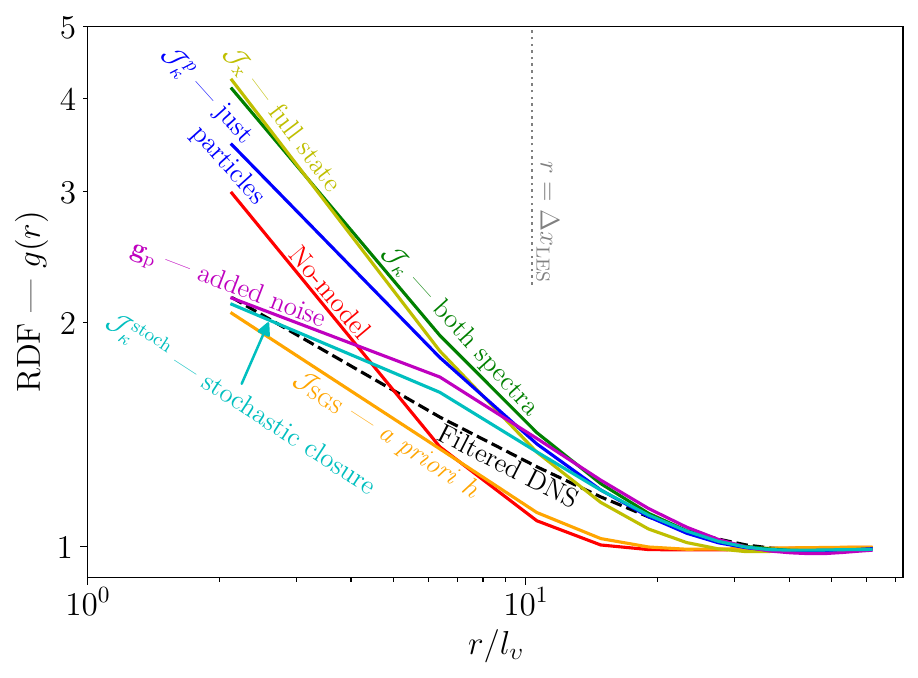}
		\caption{Time-averaged radial distribution function (\ref{eq:rdf}).}\label{fig:apriori_rdf}
\end{figure}
Preferential concentration was quantified with the radial distribution function (RDF)~\cite{sundaram_collision_1997,wang_statistical_2000}, defined as the probability density function of distances between particles: 
\begin{equation}\label{eq:rdf}
	\vec{g}(r) = \frac{\vec{n}(r)}{N_p \rho_n dA},
\end{equation}
where $\vec{n}(r)$ is the number of particle pairs separated by distance $r$, $dA=2\pi dr$ is the area of the corresponding differential annulus, and $\rho_n=N_p/L^2$ is the mean particle concentration.
The RDF was computed with $50$ bins up to a separation of $r=0.125L$.
Figure~\ref{fig:apriori_rdf} shows time-averaged RDFs for all LES cases alongside those from the DNS.\@
The DNS data shows that clustering occurs below the coarse $\Delta x_{\mathrm{LES}}$ mesh scale, which makes it challenging to predict.
This is also seen in the figure~\ref{fig:vort_contour}(a) visualization, which also shows excessive clustering in the no-model LES (figure~\ref{fig:vort_contour}(b)).
The $\mathcal{J}_\mathrm{x}$ neural-network closure is able to improve the structure somewhat by forming the long, thin particle streaks in figure~\ref{fig:vort_contour}(c).
The corresponding RDF shows that the $\mathcal{J}_\mathrm{x}$ improves preferential concentration near $r\approx \Delta x_\mathrm{LES}$ but overpredicts it for small separations, which corresponds to the narrow particle chains visualized.

Overall, these full-match $\mathcal{J}_\mathrm{x}$ target results are far from optimal.  
The model improves on the no-model case, and it is stable, but on the whole it fails to closely match the DNS.\@
As discussed in section~\ref{sec:intro}, the sub-grid-scales become effectively stochastic for coarse mesh large-eddy simulation.  
Hence, even with training is constrained by the governing equations, it cannot necessarily reproduce the actual subgrid-scale details based on the represented scales, as is likely needed to precisely track a state trajectory.  
Therefore, we next turn to statistical objectives, which show the full potential of the formulation.  
The $\mathcal{J}_\mathrm{x}$ full-data case failure is also further assessed in section~\ref{sec:decomp_loss}, where we decompose the full field into spectral energy and phase information.

\subsection{Energy spectra mismatch objective $\mathcal{J}_\kappa$}\label{sec:spec_obj}
An important aspect of the training formulation is that it affords flexibility to target quantities of interest that are aligned with the simulation objectives.
An energy spectra mismatch
\begin{align}\label{eq:spectral_loss}
	\mathcal{J}_\kappa &=  C_u\langle \Delta\vec{E}_u,\Delta\vec{E}_u \rangle_\kappa +  C_{v_p}\langle \Delta\vec{E}_p, \Delta\vec{E}_p\rangle_\kappa,
\end{align}
is used with the motivation that it is of more direct interest and also presumably less prone to overfitting since pointwise details are not targeted.
In (\ref{eq:spectral_loss}), $\Delta\vec{E}_u = \vec{E}_u(\vec{\kappa}) - \vec{E}_u^\mathrm{DNS}(\vec{\kappa})$ and $\Delta\vec{E}_p = \vec{E}_p(\vec{\kappa}) - \vec{E}_p^\mathrm{DNS}(\vec{\kappa})$ are simply the mismatch of the spectra.

Figure~\ref{fig:pointwise}(a) shows that predictions with the $\mathcal{J}_\kappa$ training objective closely match the DNS for large scales and have only a mild surplus for the smallest represented scales, which, if deemed important, could be mitigated by adjusting the weighting in (\ref{eq:spectral_inner_product}) to emphasize high wavenumbers.
Even with far less training information, $\mathcal{J}_\kappa$ yields a more accurate particle energy distribution (figure~\ref{fig:pointwise}(b)).
However, despite its success for $\vec{E}_u$ and $\vec{E}_p$, there is essentially no improvement in the RDF (figure~\ref{fig:apriori_rdf}), which will be examined and addressed in section~\ref{sec:stochastic}.

\begin{figure} 
 	\centering
 	\includegraphics[width=0.7\linewidth]{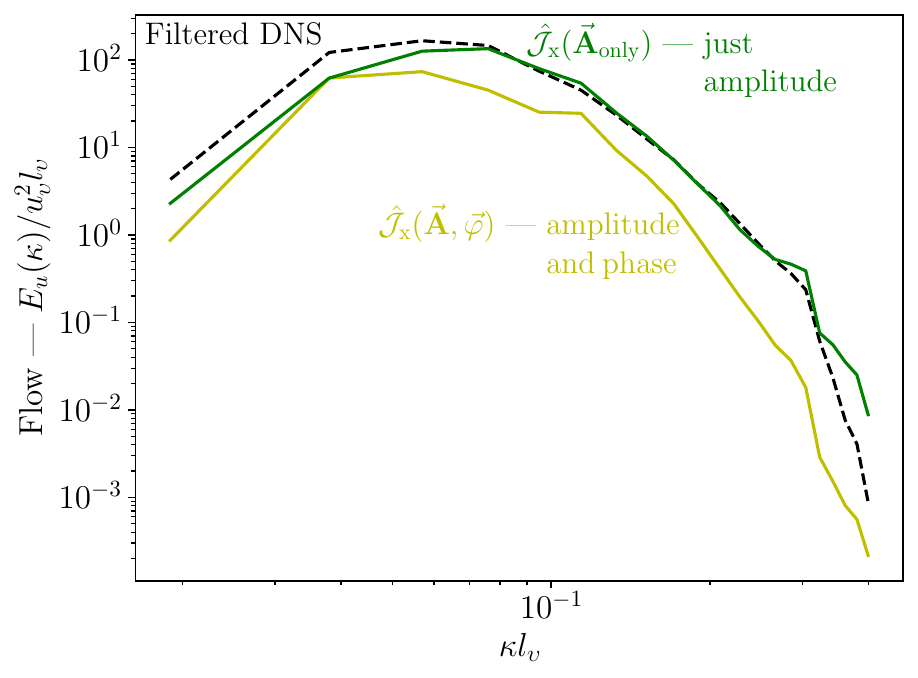}
	\captionsetup{justification=justified,singlelinecheck=false}
 	\caption{Turbulence kinetic energy spectrum normalized as in figure~\ref{fig:pointwise} trained for $\widehat{\mathcal{J}}_\mathrm{x}(\vec{\mathbf{A}},\vec{\bm{\varphi}})$ in (\ref{eq:loss_decomp}) and without phase $\widehat{\mathcal{J}}_\mathrm{x}(\vec{\mathbf{A}}_\mathrm{only})$ in (\ref{eq:loss_decomp_nophi}).}\label{fig:phases}
\end{figure}

\subsection{Amplitude and phase decomposed $\mathcal{J}_\mathrm{x}$}\label{sec:decomp_loss}

Here we analyze what about the $\mathcal{J}_\mathrm{x}$ target hinders the performance of its trained closure, relative to the much better performance for the $\mathcal{J}_\kappa$ energy spectrum target.
We design a special target to assess this:
\begin{equation}\label{eq:spec_loss_analysis}
	\widehat{\mathcal{J}}_\text{x} =  C_u\langle \widehat{\mathbf{u}}-\widehat{\mathbf{W}}, \widehat{\mathbf{u}}-\widehat{\mathbf{W}} \rangle_\kappa + C_p \langle \widehat{\mathbf{v}}_\mathrm{p} -  \widehat{\mathbf{V}}_\mathrm{p}, \widehat{\mathbf{v}}_\mathrm{p} -  \widehat{\mathbf{V}}_\mathrm{p} \rangle_\kappa,
\end{equation}
so we can separate the amplitude and phase contribution as
\begin{equation}\label{eq:loss_decomp}
\begin{split}
		\widehat{\mathcal{J}}_\text{x}(\vec{\mathbf{A}},\vec{\bm{\varphi}})  =\; & C_u\,\sum_{\vec{\bm{\kappa}}}\big[{(|\widehat{\mathbf{u}}|-|\widehat{\mathbf{W}}|)}^2 + 2|\widehat{\mathbf{u}}||\widehat{\mathbf{W}}|(1-\cos\Delta\vec{\bm{\varphi}}_{u})\big]\;+ \\
		& C_p\,\sum_{\vec{\bm{\kappa}}}\big[{(|\widehat{\mathbf{v}}_\mathrm{p}|-|\widehat{\mathbf{V}}_\mathrm{p}|)}^2 + 2|\widehat{\mathbf{v}}_{\mathrm{p}}||\widehat{\mathbf{V}}_\mathrm{p}|(1-\cos\Delta\vec{\bm{\varphi}}_{p})\big],
\end{split}
\end{equation}
where $\Delta\vec{\bm{\varphi}}_u = \vec{\bm{\varphi}}_{u} - \vec{\bm{\varphi}}_{W}$ and $\Delta\vec{\bm{\varphi}}_p = \vec{\bm{\varphi}}_{v_p} - \vec{\bm{\varphi}}_{V_p}$ are the phase differences given, for example, $\widehat{\mathbf{u}} = |\widehat{\mathbf{u}}|e^{i\vec{\bm{\varphi}}_u}$.
The first term for each flow and particle contributions in  (\ref{eq:loss_decomp}) penalizes the amplitude, as for $\mathcal{J}_\kappa$, and the second includes phase information, which is in $\mathcal{J}_\mathrm{x}$ lost in $\mathcal{J}_\kappa$.
The neural network's ability to decrease phase mismatch was assessed by training $\widehat{\mathcal{J}}_\mathrm{x}(\vec{\mathbf{A}},\vec{\bm{\varphi}})$ while holding amplitude weights---$|\widehat{\mathbf{u}}||\widehat{\mathbf{W}}|$ and $|\widehat{\mathbf{v}}_{\mathrm{p}}||\widehat{\mathbf{V}}_\mathrm{p}|$---in (\ref{eq:loss_decomp}) constant in time thereby decoupling the phase mismatch from amplitude.
Training only barely decreased the phase mismatch by 7\%.
If trained with the unmodified (\ref{eq:loss_decomp}) as in $\mathcal{J}_\mathrm{x}$, the model would apparently decrease the phase mismatch by 50\%.
This suggests that, since the $\widehat{\mathcal{J}}_\mathrm{x}(\vec{\mathbf{A}},\vec{\bm{\varphi}})$ model is unable to match phases for the smallest resolved scales, it ends up including dissipation, which reduces the phase term by reducing the amplitude. 
Training instead with
\begin{equation}\label{eq:loss_decomp_nophi}
\begin{split}
		\widehat{\mathcal{J}}_\text{x}(\vec{\mathbf{A}}_\mathrm{only}) = C_u\,\sum_{\vec{\bm{\kappa}}}{(|\widehat{\mathbf{u}}|-|\widehat{\mathbf{W}}|)}^2 + C_p\,\sum_{\vec{\bm{\kappa}}}{(|\widehat{\mathbf{v}}_{\mathrm{p}}|-|\widehat{\mathbf{V}}_{\mathrm{p}}|)}^2,
\end{split}
\end{equation}
yields far better predictions (see figure~\ref{fig:phases}).
Presumably, the full $\mathcal{J}_\mathrm{x}$ leads to a similar challenge.
This apparent importance of matching the amplitude but not phase is analogous to how non-dissipative numerical schemes (e.g., central-differences), even with potentially significant phase errors, outperform higher-order dissipative schemes~\cite{mittal_suitability_1997}.

\begin{figure}
     \begin{minipage}{0.8\textwidth} 
         \begin{subfigure}{0.5\textwidth}
             \includegraphics[width=0.9\linewidth]{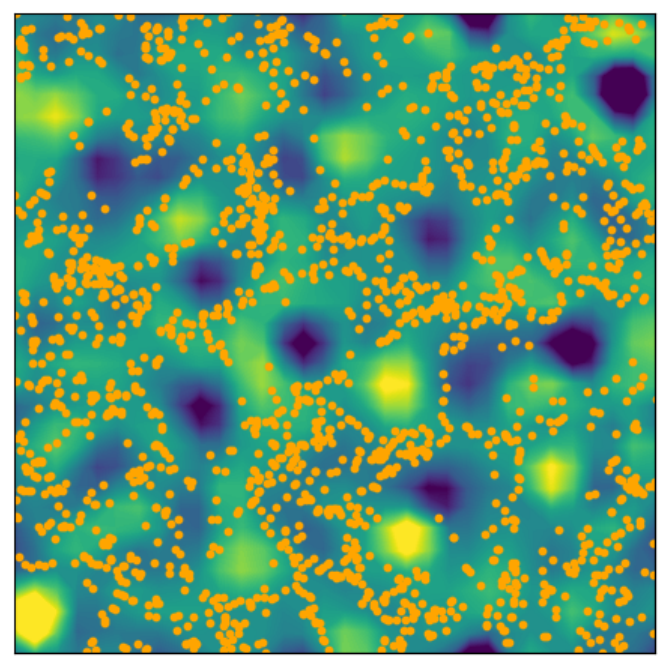}
             \caption{Filtered DNS}\label{fig:vort_cont_dns}
         \end{subfigure}%
         \begin{subfigure}{0.5\textwidth}
             \includegraphics[width=0.9\linewidth]{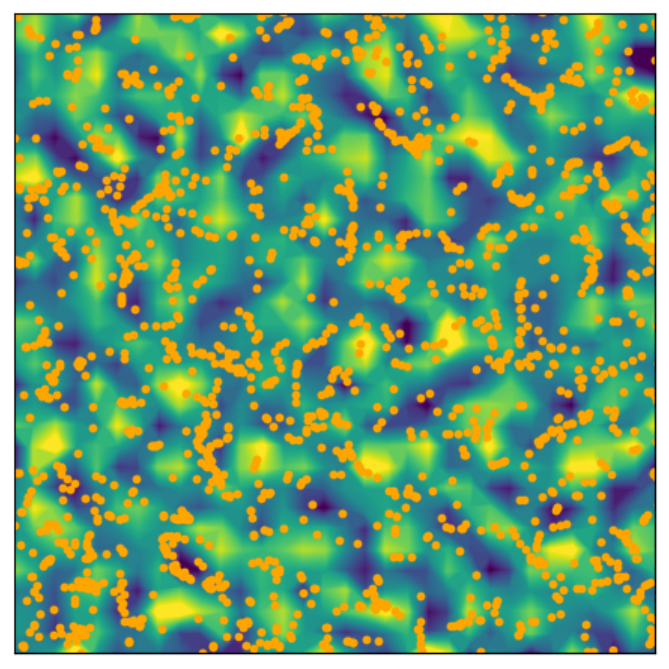}
             \caption{No-model}\label{fig:vort_cont_les}
         \end{subfigure}
         \begin{subfigure}{0.5\textwidth}
             \includegraphics[width=0.9\linewidth]{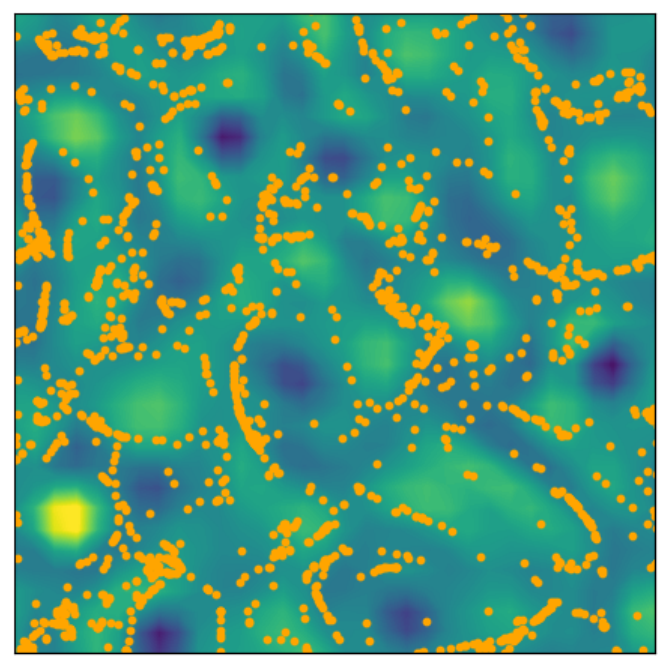}
             \caption{$\mathcal{J}_\mathrm{x}$ --- full state \\(\ref{eq:pointwise_loss})}\label{fig:vort_cont_point}
         \end{subfigure}%
		 \begin{subfigure}{0.5\textwidth}
             \includegraphics[width=0.9\linewidth]{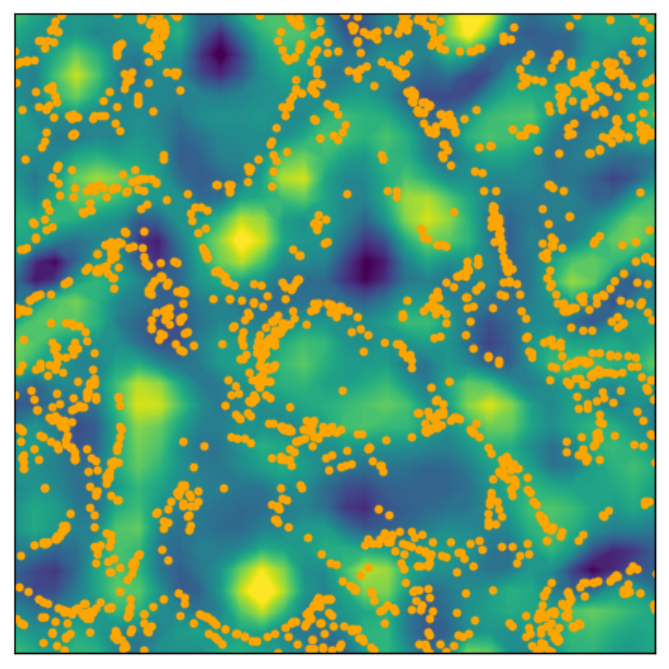}
             \caption{$\mathcal{J}_\kappa$ --- both spectra (\ref{eq:spectral_loss})}\label{fig:vort_cont_apost}
         \end{subfigure}\\
		 \begin{subfigure}{0.5\textwidth}
             \includegraphics[width=0.9\linewidth]{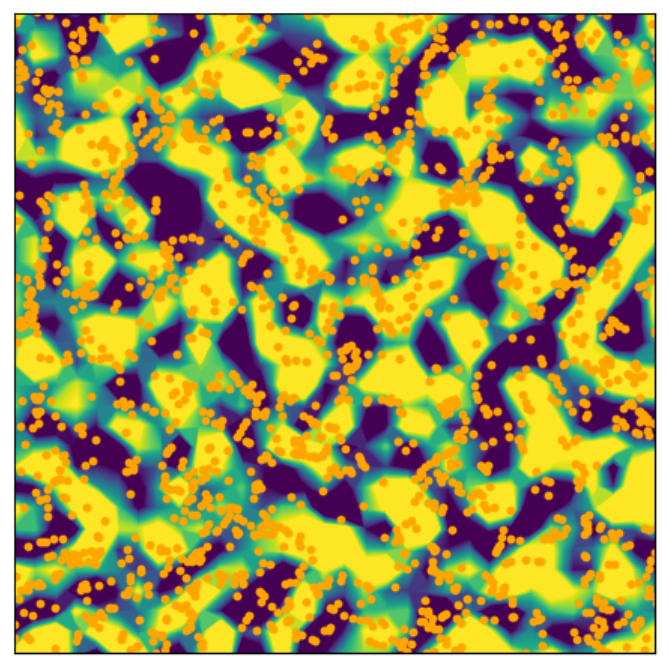}
             \caption{$\mathcal{J}_\mathrm{SGS}$ --- \textit{a priori} $h$ (\ref{eq:apiori_loss})}\label{fig:vort_cont_aprio}
         \end{subfigure}\begin{subfigure}{0.5\textwidth}
             \includegraphics[width=0.9\linewidth]{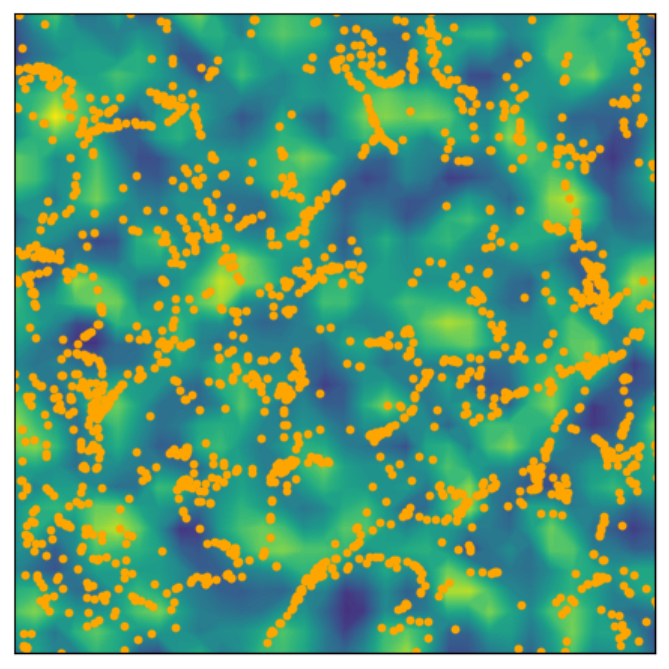}
             \caption{$\mathcal{J}_\kappa^p$ --- only particles (\ref{eq:particle_loss})}\label{fig:vort_cont_noE}
         \end{subfigure}
     \end{minipage}%
     \hspace{-0.7cm} 
     \begin{minipage}{0.2\textwidth} 
         \includegraphics[height=3.25\textwidth,trim={0 0 0 2.5mm}, clip]{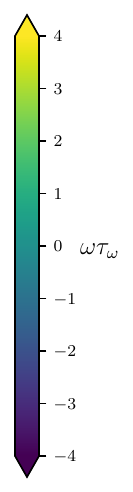} 
     \end{minipage}
	 \captionsetup{justification=justified,singlelinecheck=false}
     \caption{Vorticity visualized with color normalized by turnover time $\tau_\omega$ and particles at $t/t_\omega = 1000$. Colormap extended to avoid clipping of the \textit{a priori} $\mathcal{J}_\mathrm{SGS}$ solution.}\label{fig:vort_contour}
 \end{figure}

\subsection{Direct closure learning $\mathcal{J}_\mathrm{SGS}$}\label{sec:apriori}
With DNS data, we are able to also assess a corresponding training in which the closure is directly---and much more simply---trained to match the actual values from DNS.\@
Such information would be hard to obtain in general, as discussed in section~\ref{sec:intro}. 
It can also be anticipated that such training might not be robust when the model is used because it is not constrained by the physics of the resolved governing equations.
In this case, $\vec{\mathbf{h}}_\mathrm{u}$ and $\vec{\mathbf{h}}_\mathrm{p}$ in (\ref{eq:cont_gov_NN}) are represented by the same neural networks (\ref{eq:NN_arch}) and are trained for
\begin{equation}\label{eq:apiori_loss}
	\mathcal{J}_\text{SGS} = \langle  \Delta\vec{\bm{\tau}}, \Delta\vec{\bm{\tau}} \rangle_2+ \langle \Delta \vec{\mathbf{u}}^\prime,  \Delta\vec{\mathbf{u}}^\prime\rangle_2,
\end{equation}
where $\Delta\vec{\bm{\tau}} = \vec{\mathbf{h}}_\mathrm{u}(\vec{\bm{\phi}}_u;\vec{\theta}_u)- \vec{\bm{\tau}}^{\,r}$, $\Delta \vec{\mathbf{u}}^\prime = \vec{\mathbf{h}}_\mathrm{p}(\vec{\bm{\phi}}_p;\vec{\theta}_p) - \vec{\mathbf{u}}^\prime(\mathbf{x}_\mathrm{p})$, and $\vec{\mathbf{u}}^\prime(\vec{\mathbf{x}}_\mathrm{p}) = \mathbf{B}(\vec{\mathbf{u}}-\overline{\mathbf{u}})$ is the sub-grid velocity at the particle locations, and $\langle\cdot\rangle_2$ is the $L^2$ norm, as in (\ref{eq:inner_product}) but instantaneous in time.

It is then tested on the same out-of-sample initial conditions as in section~\ref{sec:full_obj} and section~\ref{sec:spec_obj}. 
Although training reduced $\mathcal{J}_\mathrm{SGS}$ by a factor of over $10^3$ from its initial value, the trained model yields qualitatively incorrect behavior with $k_u$ almost $25$ times and $k_p$ over $10$ times too high (see table~\ref{tab:mean_QoI}).
Overall, the results suggest that, even though $\mathcal{J}_\mathrm{SGS}$ provides a straightforward training objective, fitting it well does not guarantee accurate turbulence dynamics. 

\subsection{RDF error: discussion and correction}\label{sec:stochastic}
Figure~\ref{fig:apriori_rdf} shows that neither $\mathcal{J}_\mathrm{x}$ nor $\mathcal{J}_\kappa$ closely reproduce the RDF from the DNS.\@ 
This is probably not surprising. 
We have already confirmed that these neither closely reproduce the actual sub-grid-scale turbulence of the DNS, which is effectively stochastic, as discussed. 
Yet it is just these missing sub-grid structures that disperse the particles to effectively yield lower RDF values at small separations.  
Without them, the bands of particles in the trusted data in figure~\ref{fig:vort_contour}(a) become narrow lines of particles, such as in figure~\ref{fig:vort_contour}(d), even for the otherwise successful $\mathcal{J}_{\kappa}$ target, with a concomitant too-focused RDF.\@  
This hypothesis---a lack of stochastic sub-grid-scale fluctuations---is first assessed by adding a random perturbation $\vec{\bm{\epsilon}}\in\mathbb{R}^{N_p\times 2}$ to the $\mathcal{J}_\kappa$ trained closure,
\begin{equation}\label{eq:rand_g}
	\vec{\mathbf{g}}_\mathrm{p}(\vec{\bm{\phi}}_p;\vec{\theta}_p) = \vec{\mathbf{h}}_\mathrm{p}(\vec{\bm{\phi}}_p;\vec{\theta}_p) + \vec{\bm{\epsilon}}\odot\vec{\mathbf{h}}_\mathrm{p}(\vec{\bm{\phi}}_p;\vec{\theta}_p),
\end{equation}
where $\vec{\bm{\epsilon}}$ components are drawn from a normal distribution $\mathcal{N}(0,\sigma^2)$ each time step, similar to simple models for particle dispersion~\cite{pope_lagrangian_1994}.
The variance $\sigma^2$ was manually tuned to match the DNS RDF, which it does well as seen in figure~\ref{fig:apriori_rdf}.

Tuning can be avoided by adding an It$\mathrm{\hat{o}}$ stochastic differential equation (SDE), as commonly used in Langevin-type sub-grid-scale models~\cite{gosman_aspects_1983,minier_lagrangian_2015,pozorski_filtered_2009}:
\begin{equation}\label{eq:SDE_closure}
	d\vec{\mathbf{u}}^\prime = \vec{\mathbf{h}}_\mathrm{p}(\vec{\bm{\phi}}_p;\vec{\theta}_p)dt + \vec{\mathbf{f}}_\mathrm{p}(\vec{\bm{\phi}}_p;\vec{\theta}_p)d\vec{\mathbf{W}},
\end{equation}
where the deterministic drift $\vec{\mathbf{h}}_\mathrm{p}$ and stochastic diffusion $\vec{\mathbf{f}}_\mathrm{p}$ are learned such that the Wiener process $d\vec{\mathbf{W}}$ accounts for the additional motion induced by sub-grid-scale velocities. 
The sub-grid-scale velocity from (\ref{eq:SDE_closure}) is added to (\ref{eq:MR_unclosed}).
Equation (\ref{eq:SDE_closure}) was discretized using a stochastic fourth-order Runge--Kutta variant that is consistent with the momentum equations~\cite{Gard1988SDE}.

While the RDF well represents a goal of the training, the RDF itself is ill-suited as a training target because gradients are not well defined for the binning operations used to compute it.
Instead, the transformed number density field
\begin{equation}\label{eq:np_hat}
	\widehat{n}(\vec{\bm{\kappa}}) = \frac{1}{N_p}\sum_{j=1}^{N_p}e^{-i\vec{\mathbf{x}}_{\mathrm{p}_j}\cdot\vec{\bm{\kappa}}},
\end{equation}
is used to compute a number density fluctuation spectrum 
\begin{equation}\label{eq:en}
	\vec{E}_n(\vec{\kappa}) = \sum_{\substack{\vec{\bm{\kappa}}\in\mathcal{S}_l\\ \vec{\bm{\kappa}}\neq \mathbf{0}}}\frac{1}{N_l}|\widehat{n}(\vec{\bm{\kappa}})|^2,
\end{equation} 
for the wavenumbers $\vec{\bm{\kappa}}$ supported by the LES mesh~\cite{downing_spectral_2024,downing_spectral_2025,matsuda_influence_2014}.
Because this is sensitive to the relative proximity of particles to each other, improving it should also improve RDF predictions.

A stochastic overall objective function now includes $\vec{E}_n$ added to $\mathcal{J}_{\kappa}$ from (\ref{eq:spectral_loss}),
\begin{equation}\label{eq:stoch_loss_det}
	J_{\kappa}^\mathrm{stoch} = C_u\langle \Delta\vec{E}_u,\Delta\vec{E}_u \rangle_\kappa +  C_{v_p}\langle \Delta\vec{E}_p, \Delta\vec{E}_p\rangle_\kappa + C_{n}\langle \Delta\vec{E}_n, \Delta\vec{E}_n \rangle_{\kappa},
\end{equation}
where $\Delta\vec{E}_n = \vec{E}_n(\vec{\kappa})-\vec{E}_n^\mathrm{DNS}(\vec{\kappa})$.
For use, $J_\kappa^\mathrm{stoch}$ requires averaging to estimate $\mathcal{J}_{\kappa}^\mathrm{stoch}$. 
Hence, (\ref{eq:stoch_loss_det}) is averaged over $M$ so-called noise paths $\Delta\vec{\mathbf{W}}\in\mathbb{R}^{N_t\times N_p\times 2}$, both for its value
\begin{equation}\label{eq:stoch_loss}
	\mathcal{J}_{\kappa}^{\mathrm{stoch}}(\vec{\theta}_p) \approx \frac{1}{M}\sum_{m=1}^{M}J_{\kappa}^\mathrm{stoch}(\vec{\theta}_p;\Delta\vec{\mathbf{W}}^m),
\end{equation}
and its gradient
\begin{align}
	\nabla_{\vec{\theta}_p}\mathcal{J}_{\kappa}^{\mathrm{stoch}}(\vec{\theta}_p) &\approx \frac{1}{M}\sum_{m=1}^{M}\nabla_{\vec{\theta}_p}J_{\kappa}^\mathrm{stoch}(\vec{\theta}_p;\Delta\vec{\mathbf{W}}^m).
\end{align}
Sensitivities involving (\ref{eq:SDE_closure}) were obtained via adjoints by using the same noise paths $\Delta\vec{\mathbf{W}}^m$ that had been randomly selected for each forward solve, which provides consistent adjoint-based training despite the added noise~\cite{JMLR:v21:19-346}.

Figure~\ref{fig:apriori_rdf} shows that the resulting model improved the RDF beyond the \textit{ad hoc} (\ref{eq:rand_g}), and it also is a general example of how a stochastic model can be included in the gradient based training.
Particle distribution can be further prioritized by increasing $C_n$ in (\ref{eq:stoch_loss_det}), which prioritizes $\vec{E}_n$ when training, but yields inaccurate energy spectra.

\subsection{Limited and noisy data: toward training with experimental data}\label{sec:robustness}
DNS conveniently provides data for developing and testing closures, but closures are most useful in regimes where DNS is not possible, either for its expense or that the physics is incompletely described in the governing equations.
Experimental measurements are therefore attractive, though these rarely provide the full two-phase state information.
They also necessarily involve some errors or uncertainties, often manifest as measurement noise. 
We next assess how robust training is in such cases by restricting and corrupting the training data.

\subsubsection{Particle-only data}\label{sec:part_obj}

For many experiments---especially at high particle loading---flow measurements are more difficult than particle tracking.
Accordingly, we attempt to train only on particle statistics with
\begin{equation}\label{eq:particle_loss}
 \mathcal{J}^p_\kappa =  \langle \Delta\vec{E}_p, \Delta\vec{E}_p \rangle_\kappa,
\end{equation}
which includes only the particle contribution to the successful $\mathcal{J}_\kappa$ in (\ref{eq:spectral_loss}).
Spectra in figure~\ref{fig:pointwise} and RDF in figure~\ref{fig:apriori_rdf} show that $\mathcal{J}^p_\kappa$ yields predictions that are nearly as good as the full $\mathcal{J}_\kappa$ training (and much better than $\mathcal{J}_\mathrm{x}$; see table~\ref{tab:mean_QoI}). 
This suggests that the missing flow information is sufficiently encoded in the inertial particle statistics, consistent with Csanady’s transfer function, which relates inertial-particle measurements to the corresponding tracer (Lagrangian) spectrum~\cite{csanady_turbulent_1963}.
Inferring the missing information leverages the constraint of the governing equations in the training.
Higher Stokes number particles increasingly decorrelate from the small flow scales, so learning a closure from heavier particles will be more challenging, though $St=8$ is still nearly as successful (not shown).

\subsubsection{Noisy particle velocities}\label{sec:noise}

\begin{figure}
	\begin{subfigure}{0.8\textwidth}
		\includegraphics[width=\linewidth]{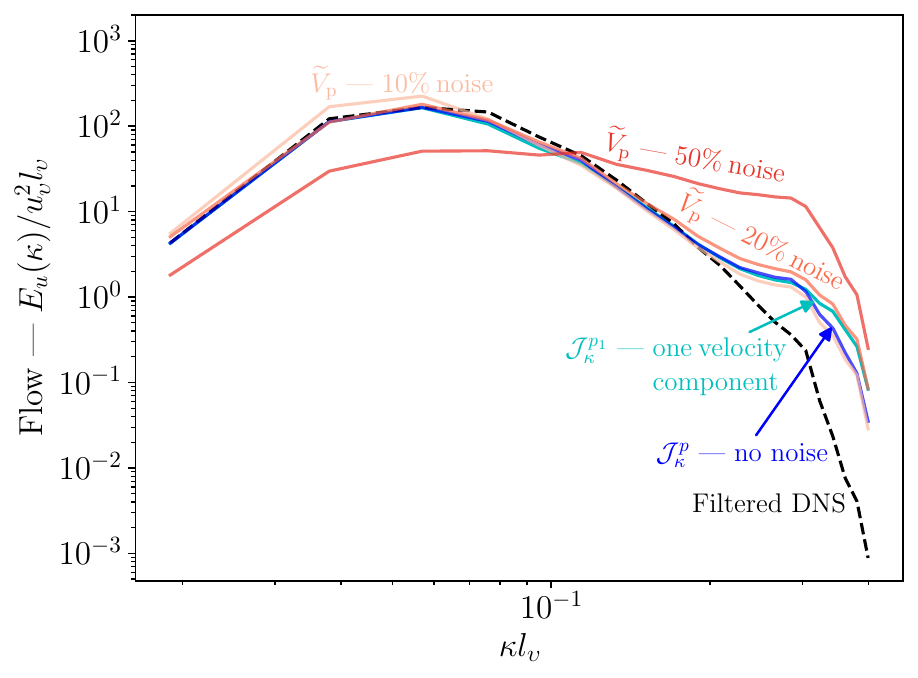}
		\caption{Turbulence kinetic energy}\label{fig:tke_noise}
	\end{subfigure}
	\begin{subfigure}{0.8\textwidth}
		\includegraphics[width=\linewidth]{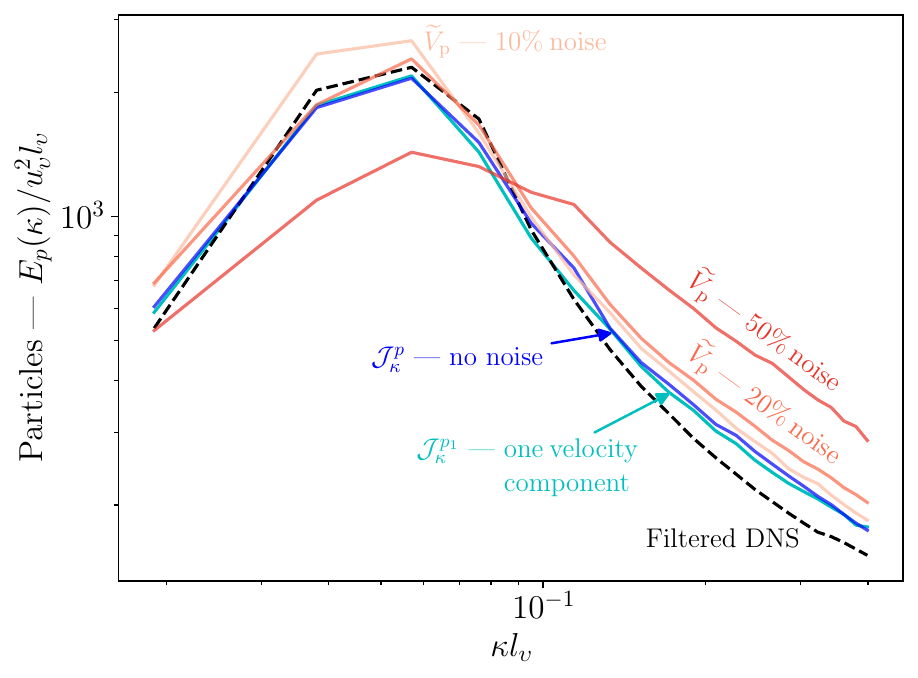}
		\caption{Particle kinetic energy}\label{fig:vpke_noise}
	\end{subfigure}
	\captionsetup{justification=justified,singlelinecheck=false}
	\caption{Time-averaged spectra trained with noisy data, only one velocity component, and the particle spectra $\mathcal{J}^p_\kappa$ (\ref{eq:particle_loss}) loss function for reference.}\label{fig:noise}
\end{figure}
In this case, measurement noise was modeled simply by perturbing the DNS particle locations $\vec{\mathbf{X}}_p$ and velocities $\vec{\mathbf{V}}_p$ used in $\mathcal{J}_\kappa$ (\ref{eq:spectral_loss}):
\begin{subequations}
	\begin{align}
		\widetilde{\mathbf{X}}_p &= \vec{\mathbf{X}}_p + \vec{\bm{\epsilon}}_x,\\
		\widetilde{\mathbf{V}}_p &= \vec{\mathbf{V}}_p + \vec{\bm{\epsilon}}_v,
	\end{align}
\end{subequations}
where $\vec{\bm{\epsilon}}_x, \vec{\bm{\epsilon}}_v\in\mathbb{R}^{N_p\times 2}$.
Perturbations $\vec{\bm{\epsilon}}_x$ and $\vec{\bm{\epsilon}}_v$ components are sampled each time step from a normal distribution $\vec{\bm{\epsilon}}_x\sim\mathcal{N}(0;\sigma_x^2)$ and $\vec{\bm{\epsilon}}_v\sim\mathcal{N}(0;\sigma_v^2)$, respectively.
This resembles experimental uncertainty in particle positions, which lead to uncertainty in particle displacements and therefore velocities.
The particle perturbation $\vec{\bm{\epsilon}}_x$ was adjusted to achieve $\sigma_v = 10\%$, $20\%$, and $50\%$ of $V_{p_\mathrm{rms}}$.
The standard deviation from $\vec{\bm{\epsilon}}_x$ and $\vec{\bm{\epsilon}}_v$ can be related using first-order finite-differences: $\sigma_v = C_\sigma \sigma_x/\Delta t$, which yields $\sigma_x \approx 1\%$, $2\%$, and $5\%$ of the integral length scale $l_k$.
The model was then trained just as in section~\ref{sec:spec_obj}.

The predicted energy spectra in figure~\ref{fig:noise} are remarkably robust to this noise.
For $10\%$ and $20\%$ noise the flow and particle spectra are still well predicted.
They only start to become qualitatively different for $\sigma_v \gtrsim 0.2V_{p_\mathrm{rms}}$, where the noise prevents the NN from learning a useful closure, hence reverting to a flat spectrum similar to the no-model LES in figure~\ref{fig:pointwise}. 
Nonetheless, this demonstrates the robustness conferred by PDE constraints in training.

\subsubsection{Missing velocity components}\label{sec:one_vp_comp}

Measurement techniques resolving individual particle kinematics, such as particle tracking velocimetry, most often provide planar information for low to moderate particle loading.
We emulate this limited observability by training using information from only the first of the two particle-velocity components. 
The single-component particle-velocity spectrum is
\begin{equation}\label{eq:ep_i}
	\vec{E}_{p_1}^\circ(\vec{\kappa}) = \sum_{\vec{\bm{\kappa}}\in\mathcal{S}_l}\frac{1}{N_l}\left|\widehat{v}_{p_1}(\vec{\bm{\kappa}})\right|^2,
\end{equation}
which is rescaled as before for energy consistency,
\begin{equation}\label{eq:ep_rescaled_i}
	\vec{E}_{p_1}(\vec{\kappa}) = 
	\frac{\langle v_{p_1}^2\rangle_p}{\sum_{\vec{\kappa}}\vec{E}_{p_1}^\circ(\vec{\kappa})}\,\vec{E}_{p_1}^\circ(\vec{\kappa}).
\end{equation}
The objective is
\begin{equation}\label{eq:particle_loss_1}
 \mathcal{J}^{p_1}_\kappa =  \langle \Delta\vec{E}_{p_1}, \Delta\vec{E}_{p_1} \rangle_\kappa,
\end{equation}
where $\Delta\vec{E}_{p_1} = \vec{E}_{p_1} - \vec{E}_{p_1}^\mathrm{DNS}$.
Because only one velocity component is provided, the training objective does not supply the model with isotropy (or any assumed relationship between components), yet figure~\ref{fig:noise} shows that the prediction is nearly as good.
We compute a normal-stress isotropy deviation 
\begin{equation}\label{eq:R_crit}
	\mathcal{R} = \frac{|R_{11}-R_{22}|}{\frac{1}{2}(R_{11}+R_{22})},
\end{equation}
from the Reynolds-stress tensor $R_{ij} = \langle \vec{u}_i \vec{u}_j\rangle_{\mathrm{x}}$.
Even for a single particle‑velocity component it does not show a significant deviation from isotropy, $\mathcal{R}^{p_1} = 0.035$, and it is nearly the same as using both the flow and the particle spectra with $\mathcal{J}_\kappa$ ($\mathcal{R}^{p} = 0.032$).
Incorporating isotropy as a training requirement may improve the flow symmetry.

\begin{table}
	\centering
	\captionsetup{justification=justified,singlelinecheck=false}
	\begin{tabular}{cccc}\hline
	Particles used & \:$n_p l_k^2/L^2$ & \,$k_u/k_{u}^{\mathrm{DNS}}$ &\,$k_p/k_{p}^\mathrm{DNS}$ \\ \hline\hline
	$100\%$ & 13 & 0.95 & 1.03 \\
	$50\%$  & 6 & 0.93 & 0.96 \\
	$25\%$  & 3 & 0.86 & 0.70 \\
	$10\%$  & 1 & 1.50 & 0.94 \\ \hline
	\end{tabular}
	\caption{Kinetic energy from LES with neural-network closures trained on varying percentage of total particles.}\label{tab:k_vs_Np}
\end{table}

\subsubsection{Incomplete particle data}

It is challenging to track all particles in any experiment, and we can anticipate that training effectiveness will depend on the available sample.
We assess this by training closures for $\mathcal{J}_{\kappa}^p$ using a randomly selected subset $n_p$ of the available particles $N_p$.
Since the closure prediction varied according to the subset of particles used in training, the neural-network weights were initialized, trained, and evaluated $30$ times with a new set of random $n_p$ each time.
Table~\ref{tab:k_vs_Np} reports the ensemble-averaged turbulence and particle kinetic energies as a function of the average number of particles within an energy-containing eddy of area $\sim l_{k}^2$.
The all-particle results reported in section~\ref{sec:spec_obj} have approximately $13$ particles per eddy.
Using half the number of particles only increased the prediction error by $3\%$ for $k_u$ and $5\%$ for $k_p$.
For sparse sampling---less than 3 particles per nominal eddy---the closure incurs $\gtrsim 14\%$ error.

\section{Conclusions and additional discussion}\label{sec:conc}

The primary conclusion is that incorporating the full discretized governing equations as a constraint when forming gradients for training provides remarkable benefits. Foremost, it yields robust and accurate models. Though far simpler, regression training on the closure, which in truth would not itself be available in applications, does not yield a viable model (even though it does provide a seemingly reasonable fit of the closure). The constrained formulation is also robust in the sense that a subset of the data is sufficient. We demonstrated this with training targets based on just particles for training sub-grid-scale turbulence stresses, which only slightly diminishes accuracy. This is still effective when using a subset of available particles, particles with added measurement-like noise, and even including just one velocity component. The main reason for this capacity to extrapolate is that the training is constrained by the full coupled physics of the discretized governing equations.

As with any model, there are also limitations. Selection of training targets required attention. Unexpectedly, using the full available trusted data actually impedes training. It seemed that this is a manifestation of a known property of sub-grid-scale models relative to high-resolution fields: there is no unique sub-grid realization corresponding to a coarse representation of turbulence. Reproducing the sub-grid scales from a direct numerical simulation is not necessarily best for a corresponding large-eddy simulation. A special target function we developed suggested in particular that enforcing phase information, as opposed to amplitude, hindered the training. This is similar to how non-dissipative methods, even though they might have significant phase errors, are advantageous for turbulence because they preserve the energy spectrum.  

Another, though indirect, consequence of the unrecoverable sub-grid-scale turbulence details was seen in the particle dynamics: the accurate sub-grid models did not provide the full particle dispersion of the direct numerical simulation. Particle spectra were well predicted, but RDFs were too sharp and visualizations showed narrow trains of particles that should be more diffuse. Such behavior is a consequence of employing a deterministic model on so coarse a mesh. Adding a simple stochastic term supported this by recovering the dispersion, though it required hand tuning. More importantly, we were able to train the coefficients of a stochastic model using the same procedure, eliminating the need for tuning. This also demonstrated the flexibility of the approach for more general model formulations.

This study focused on two dimensions to facilitate the exploration of many cases. Still, it includes the key mechanisms of turbulence-particle suspension in large-eddy simulation: sub-grid-scale turbulence stresses and two-way particle coupling with both represented and unrepresented scales. The method has already been demonstrated on single-phase flow turbulence~\cite{macart_embedded_2021,sirignano_dpm_2020}.

Of course, the entire procedure depends on the differentiation of the model-embedded governing equations, which can be onerous to construct by hand. In the implementation used for this study, the discrete adjoints were hand-constructed, which is relatively straightforward for the staggered-mesh methods and periodic boundaries. Inclusion of particles was facilitated by well-defined operators such as $\mathbf{B}$ and $\mathbf{B}^T$ in (\ref{eq:disc_gov}) for coupling flow and the particles. Automatic differentiation is another option for doing this, and it is particularly attractive for more intricate systems, so long as it can provide sufficiently efficient code. Many machine-learning engines support automatic differentiation, since it is routinely used to compute gradients of the neural network itself~\cite{baydin_automatic_2018}. We can anticipate that training will generally converge most quickly when the target functions are continuously differentiable, which is why the RDF was not directly targeted in section~\ref{sec:stochastic}. Without smoothness, we can anticipate slower or even unsuccessful training, though convergence might not be necessary to significantly improve a model. The success of the model is not the same as a converged training regimen.

Our most successful training objectives were based on spatial spectra; however there are many other targets known to provide similar information, such as two-point correlations or wavelet-based band energies. By Parzen windowing the domain~\cite{NumericalRecipes}, as we might for a non-periodic domain, we intentionally break the periodicity when training. This yields spectra nearly as accurate as the standard $\mathcal{J}_\kappa$ case. By including additional DNS data processed with different filter/grid sizes during training, as described in section~\ref{sec:training_data}, we were able to train a successful model for $N^2 = 24^2$, $32^2$, and $48^2$ with no changes to the neural network architecture. Models trained this way performed slightly better than trained on a single resolution. As one might expect, more aggressive filtering makes training a closure for coarser resolutions more challenging. We also attempted to generalize a model for $St = 0.8$ and $8$ in the same way. Models trained with both $St$ simultaneously still outperformed the no-model LES, but underperformed ones trained and tested on a single $St$. Significant improvements were obtained by using feature-wise
linear modulations~\cite{perez2018film} in the neural-network architecture, which allowed models trained on both $St$ simultaneously to match the performance of those trained on a single $St$. We have not yet attempted to train across $Re$ and volume loading $\Phi_v$. With the base approach established, further testing will be needed to examine this. In general, the robustness makes the approach seem well-suited for incorporating additional physics, such as phase change, particle--particle collisions, or reaction, all of which will benefit from the physical constraints afforded by the approach.

A potential challenge is the flow evolution time needed for training. The chaotic amplification of perturbations in turbulence makes adjoint solutions unstable. Despite that, the general success of the method for this application confirms that our adjoint solves, as discussed in section~\ref{sec:training}, are for times sufficient to achieve the training objective. The necessary relational time scale between the inputs (e.g., one component of the velocities) and the actions that achieve the objective (e.g., the flow sub-grid-scale stress) can therefore be concluded to be represented in the training time window. This is expected to depend on the objective, and it might warrant more consideration in some cases. In our tests, the spectrum-based objective $\mathcal{J}_{\kappa}$ did not make the adjoint dynamics qualitatively more stable; it exhibited comparable blow-up rates as the pointwise objective $\mathcal{J}_\mathrm{x}$. If needed, more advanced optimization procedures designed to skirt the impact of the chaos in forming gradients might be useful~\cite{BLONIGAN2017803,CHUNG2022111077,WANG2014210}. Nevertheless, training was successful without special considerations for this application.  

Finally, we note that these procedures can be viewed, at least informally, from an information~\cite{Duran_Information_2022} or causality~\cite{Lozano-Duran_Bae_Encinar_2020,Tissot_2014} perspective. One interpretation of the success of the sub-grid-scale model trained on just one component of the particle velocity spectrum is that, when constrained by the governing equations, it surprisingly includes sufficient information to inform the consequential dynamics leading to the sub-grid-scale models. This is potentially helpful for diagnosing the workings of particle-laden turbulence, and it could extend to still more complex flow systems. Machine learning methods have been proposed for filling in missing data~\cite{aksoy_reconstruction_2023,morimoto_experimental_2021}; optimizations such as we employ---with the constraint of the full governing equations---allow extension to physics well beyond the direct measurements. Generalizing this to other conditions with additional physical interactions might expand the realm of what can be measured---indirectly via inference---with experimental diagnostics. A single velocity component of suspended, two-way-coupled particles can be sufficient to infer consistent sub-grid-scale stresses. Any adjoint solution provides a sensitive field; going further to train models with them establishes a firmer connection based on consequential information flow. It is consequential information in the sense that it can improve a model. Using these methods in this way is an intriguing direction for future development, a step beyond finding a useful closure for prediction.

\appendix

\section{Discretization details}\label{app:discret_det}
This appendix contains additional information regarding the discretization.
\subsection{Spatial discretization}
A uniform staggered mesh was used with the offset $\Delta$. 
The velocities were located at 
\begin{equation*}
	\vec{\mathbf{x}} =
	\begin{bmatrix}
		(i\Delta,j\Delta+\frac{\Delta}{2})\\
		(i\Delta+\frac{\Delta}{2},j\Delta) 
	\end{bmatrix},\quad i,j = 0,\ldots,N-1,
\end{equation*}
where $N$ is the number of mesh points in each direction, and pressure was located at the cell centers $(i\Delta+\frac{\Delta}{2},j\Delta+\frac{\Delta}{2})$.
The gradient, divergence, and interpolation operators, $\mathbf{G}={[G_1,G_2]}^T$, $\mathbf{D}=-[G_1^\top,G_2^\top]$, and $\mathbf{J}={[J_1,J_2]}^T$ are defined as:
\begin{align}
& G_1 = G\otimes I, \quad J_1= J\otimes I, \\
& G_2 = I\otimes G, \quad J_2 = I\otimes J,
\end{align}
where $I$ is the identity matrix, $\otimes$ is the discrete outer product, and
\begin{align}
	& G = \frac{1}{\Delta}\text{tridiag}[0,-1,1],\\
	& J = \text{tridiag}[1/2,1/2,0],
\end{align}
are cyclic tridiagonal matrices for our periodic domain. 
The Laplace operator is defined $L = -{G_1}^T G_1-{G_2}^T G_2\in\mathbb{R}^{N^2\times N^2}$.
The particle interpolation operators $\mathbf{B}={[B_1, B_2]}^T\in\mathbb{R}^{ N_p\times 2\times N^2}$ are sparse matrices with each row containing the corresponding linear interpolation weights for the corresponding particle. 
Thus, the governing equations (\ref{eq:cont_gov_NN}) are discretized in space as 
\begin{subequations}
	\begin{align}
		\mathbf{D} \vec{\mathbf{u}} & = 0,  \\ 
		\frac{d \vec{\mathbf{u}}}{dt} +\mathbf{D} [{(\mathbf{J}\vec{\mathbf{u}}^T)}^T(\mathbf{J} \vec{\mathbf{u}}^T)] & = -\mathbf{G}\vec{p} - \upsilon L^{2}\vec{\mathbf{u}}-\alpha L^{-4}\vec{\mathbf{u}}- \vec{\mathbf{F}}_\mathrm{p}+\vec{\mathbf{F}}_{\mathrm{u}}+ \mathbf{D}\vec{\mathbf{h}}_\mathrm{u}(\vec{\bm{\phi}}_u;\vec{\theta}_{u}) , \label{eq:semidisc_NS}\\
		\frac{d \vec{\mathbf{v}}_\mathrm{p}}{dt} & = \frac{1}{\tau_p}(\mathbf{B}\vec{\mathbf{u}}-\vec{\mathbf{v}}_\mathrm{p})-\vec{\mathbf{h}}_\mathrm{p}(\vec{\bm{\phi}}_p; \vec{\theta}_{p}), \\ 
		 \frac{d \vec{\mathbf{x}}_\mathrm{p}}{dt}&=\vec{\mathbf{v}}_\mathrm{p},
	\end{align}
\end{subequations}
where $\vec{\mathbf{u}}\in\mathbb{R}^{N^2\times 2}$ is a list of the flow velocity at $N^2$ meshpoints, and the particle velocity and position are $\vec{\mathbf{v}}_\mathrm{p},\vec{\mathbf{x}}_\mathrm{p}\in\mathbb{R}^{N_p\times 2}$ for $N_p$ particles.
The coupling term in (\ref{eq:semidisc_NS}) is
	\begin{equation}
		\vec{\mathbf{F}}_\mathrm{p} =\frac{m_p}{m_f}\mathbf{B}^T\frac{d\vec{\mathbf{v}}_\mathrm{p}}{dt},
	\end{equation} 
where $\mathbf{B}^T$ is the corresponding particle-to-face interpolation operator~\cite{prosperetti_computational_2007}.
\subsection{Incompressibility constraint}\label{app:projection}
A projection operation was used to correct the velocity field every Runge--Kutta sub-step, 
\begin{subequations}\label{eq:fractionalStep}
	\begin{align}
		L\vec{p}^{\: n,s} &= \frac{1}{\Delta t}\mathbf{D}\tilde{\mathbf{u}}^{n,s},\\
		\vec{\mathbf{u}}^{n,s} &= \tilde{\mathbf{u}}^{n,s} - \Delta t\mathbf{G}\vec{p}^{\:n,s},
	\end{align}
\end{subequations}
where $\tilde{\mathbf{u}}^{n,s}$ is the divergent velocity field (see section~\ref{sec:temp_disc}). 
We can write (\ref{eq:fractionalStep}) without the nominal pressure as 
\begin{equation}
	\vec{\mathbf{u}}^{n,s} = \tilde{\mathbf{u}}^{n,s}-\mathbf{G}L^{-1}\mathbf{D} \tilde{\mathbf{u}}^{n,s},
\end{equation}
where $L^{-1}$ is the inverse Laplace operator. 
Our Poisson solver uses a second-order spectral finite difference, consistent with other spatial operators. 
This can be expressed concisely as 
\begin{equation}
	\vec{\mathbf{u}}^{n,s} = \mathbf{P}\tilde{\mathbf{u}}^{n,s},
\end{equation}
where $\mathbf{P} = \bm{\delta}-\mathbf{G}L^{-1}\mathbf{D}$ and $\bm{\delta}$ is the Kronecker delta. 
This notation will be used in the following sections.
\subsection{Temporal discretization}\label{sec:temp_disc}
The four-stage Runge--Kutta with projection step in its standard form is\begin{subequations}\label{eq:rk4_NS}
	\begin{align}
		\tilde{\mathbf{u}} ^{n,1} =\;& \vec{\mathbf{u}} ^{n-1,4} + \frac{\Delta t}{2}\vec{\mathbf{R}}_\mathrm{u}^{n-1,4}, \\
		 \vec{\mathbf{u}}^{n,1} =\;& \mathbf{P}\tilde{\mathbf{u}}^{n,1},\\
		\tilde{\mathbf{u}} ^{n,2}  =\;& \vec{\mathbf{u}} ^{n-1,4} + \frac{\Delta t}{2}\vec{\mathbf{R}}_\mathrm{u}^{n,1}, \\
		 \vec{\mathbf{u}}^{n,2} =\;& \mathbf{P}\tilde{\mathbf{u}}^{n,2},\\
		\tilde{\mathbf{u}} ^{n,3} =\;& \vec{\mathbf{u}} ^{n-1,4} + \Delta t\vec{\mathbf{R}}_\mathrm{u}^{n,2}, \\
		 \vec{\mathbf{u}}^{n,3} =\;& \mathbf{P}\tilde{\mathbf{u}}^{n,3},\\
		\tilde{\mathbf{u}} ^{n,4} =\;&  \vec{\mathbf{u}} ^{n-1,4} + \frac{\Delta t}{6}(\vec{\mathbf{R}}_\mathrm{u}^{n-1,4}  + 2\vec{\mathbf{R}}_\mathrm{u}^{n,1} + 2\vec{\mathbf{R}}_\mathrm{u}^{n,2}+  \vec{\mathbf{R}}_\mathrm{u}^{n,3}), \\
		 \vec{\mathbf{u}}^{n,4} =\;& \mathbf{P}\tilde{\mathbf{u}}^{n,4},
	\end{align}
\end{subequations}
where superscripts $n$ and $s$ indicate time step and sub-step, $\vec{\mathbf{R}}_\mathrm{u}^{n,s} = -\mathbf{D} [{(\mathbf{J}{(\vec{\mathbf{u}}^{n,s})}^T)}^T\odot\mathbf{J} {(\vec{\mathbf{u}}^{n,s})}^T] - \upsilon L^{2}\vec{\mathbf{u}}^{n,s} -\alpha L^{-4}\vec{\mathbf{u}}^{n,s}- \vec{\mathbf{F}}_\mathrm{p}^{n,s}+\vec{\mathbf{F}}_\mathrm{u}^{n,s} + \mathbf{D}\vec{\mathbf{h}}_\mathrm{u}(\vec{\bm{\phi}}_u^{n,s};\vec{\theta}_{u})$. 
In this notation, for $n>1$, $\vec{\mathbf{u}} ^{n,4}$ is the approximation of $\vec{\mathbf{u}}(t_n)$, and $\vec{\mathbf{u}} ^{0,4}$ is the initial condition. 
We rewrite this in a more compact form that will be useful for the adjoint derivation in appendix~\ref{app: adj_der}:
\begin{subequations}\label{eq:discrete_NS}
	\begin{align}
		\vec{\mathbf{N}}^{n,1}  =\;& \frac{2\vec{\mathbf{u}}^{n,1}-2\mathbf{P}\vec{\mathbf{u}} ^{n-1,4}}{\Delta t}  - \mathbf{P}\vec{\mathbf{R}}_\mathrm{u}^{n-1,4},\\
		\vec{\mathbf{N}}^{n,2} =\;& \frac{2\vec{\mathbf{u}} ^{n,2}-2\mathbf{P}\vec{\mathbf{u}} ^{n-1,4}}{\Delta t}  -  \mathbf{P}\vec{\mathbf{R}}_\mathrm{u}^{n,1}, \\
		\vec{\mathbf{N}}^{n,3} =\;& \frac{\vec{\mathbf{u}} ^{n,3}-\mathbf{P}\vec{\mathbf{u}} ^{n-1,4}}{\Delta t}  - \mathbf{P}\vec{\mathbf{R}}_\mathrm{u}^{n,2}, \\
		\vec{\mathbf{N}}^{n,4} =\;& \frac{6\vec{\mathbf{u}} ^{n,4}-2\mathbf{P}(\vec{\mathbf{u}} ^{n,1}+2\vec{\mathbf{u}} ^{n,2}+\vec{\mathbf{u}} ^{n,3}-\vec{\mathbf{u}} ^{n-1,4})}{\Delta t} -  \mathbf{P}\vec{\mathbf{R}}_\mathrm{u}^{n,3}.
	\end{align}
\end{subequations}
Similarly, the discrete Maxey--Riley--Gautignol and corresponding kinematics equations can be written as
\begin{subequations}\label{eq:discrete_MR}
	\begin{align}
		\vec{\mathbf{M}}^{n,1} =\;& \frac{2\vec{\mathbf{v}}_\mathrm{p} ^{n,1}-2\vec{\mathbf{v}}_\mathrm{p}^{n-1,4}}{\Delta t}  - \vec{\mathbf{R}}_\mathrm{v_p}^{n-1,4}, \\
		\vec{\mathbf{M}}^{n,2} =\;& \frac{2\vec{\mathbf{v}}_\mathrm{p} ^{n,2}-2\vec{\mathbf{v}}_\mathrm{p}^{n-1,4}}{\Delta t}  -  \vec{\mathbf{R}}_\mathrm{v_p}^{n,1}, \\
		\vec{\mathbf{M}}^{n,3} = \;&\frac{\vec{\mathbf{v}}_\mathrm{p} ^{n,3}-\vec{\mathbf{v}}_\mathrm{p}^{n-1,4}}{\Delta t}  -\vec{\mathbf{R}}_\mathrm{v_p}^{n,2},\\
		\vec{\mathbf{M}}^{n,4}=\;& \frac{6\vec{\mathbf{v}}_\mathrm{p} ^{n,4}-2(\vec{\mathbf{v}}_\mathrm{p}^{n,1}+2\vec{\mathbf{v}}_\mathrm{p} ^{n,2}+\vec{\mathbf{v}}_\mathrm{p} ^{n,3}-\vec{\mathbf{v}}_\mathrm{p}^{n-1,4})}{\Delta t} - \vec{\mathbf{R}}_\mathrm{v_p}^{n,3},
	\end{align}
\end{subequations} and \begin{subequations}\label{eq:discrete_xp}
	\begin{align}
		\vec{\mathbf{X}}^{n,1} =\;&\frac{2\vec{\mathbf{x}}_\mathrm{p} ^{n,1}-2\vec{\mathbf{x}}_\mathrm{p}^{n-1,4}}{\Delta t}  - \vec{\mathbf{R}}_\mathrm{x_p}^{n-1,4},\\
		\vec{\mathbf{X}}^{n,2} =\;& \frac{2\vec{\mathbf{x}}_\mathrm{p} ^{n,2}-2\vec{\mathbf{x}}_\mathrm{p}^{n-1,4}}{\Delta t}  -  \vec{\mathbf{R}}_\mathrm{x_p}^{n,1},\\
		\vec{\mathbf{X}}^{n,3} =\;& \frac{\vec{\mathbf{x}}_\mathrm{p} ^{n,3}-\vec{\mathbf{x}}_\mathrm{p}^{n-1,4}}{\Delta t}  -\vec{\mathbf{R}}_\mathrm{x_p}^{n,2}, \\
		\vec{\mathbf{X}}^{n,4} =\;& \frac{6\vec{\mathbf{x}}_\mathrm{p} ^{n,4}-2(\vec{\mathbf{x}}_\mathrm{p}^{n,1}+2\vec{\mathbf{x}}_\mathrm{p} ^{n,2}+\vec{\mathbf{x}}_\mathrm{p} ^{n,3}-\vec{\mathbf{x}}_\mathrm{p} ^{n-1,4})}{\Delta t} - \vec{\mathbf{R}}_\mathrm{x_p}^{n,3},
	\end{align}
\end{subequations}
where $\vec{\mathbf{R}}_\mathrm{v_p}^{n,s} = \frac{1}{\tau_p}(\mathbf{B}\vec{\mathbf{u}}^{n,s}-\vec{\mathbf{v}}_\mathrm{p}^{n,s})+\vec{\mathbf{h}}_\mathrm{p}(\vec{\bm{\phi}}_p^{n,s};\vec{\theta}_{p})$ and  $\vec{\mathbf{R}}_\mathrm{x_p}^{n,s}=\vec{\mathbf{v}}_\mathrm{p}^{n,s}$.

\section{Adjoint equations}\label{app: adj_der}
 The adjoint equations can be obtained by expanding the time series in each of the left-hand-side terms in (\ref{eq:u_adj_compact}), (\ref{eq:vp_adj_compact}), and (\ref{eq:xp_adj_compact}). 
 Terms at the $n^{\text{th}}$ and $s^{\text{th}}$ time level are then gathered to obtain the  Runge--Kutta form of the equations. 
 This process is omitted for conciseness. Once completed, it yields 
\begin{subequations}\label{eq:adj_u}
	\begin{align}
		\vec{\mathbf{N}}^{\dagger n+1,3} =\;& \frac{\mathbf{P}\vec{\mathbf{u}} ^{\dagger n+1,3}-\vec{\mathbf{u}} ^{\dagger n+1,4}}{\Delta t}  - \frac{1}{2}\vec{\mathbf{R}}_\mathrm{u}^{\dagger n+1,3}, \\
		\vec{\mathbf{N}}^{\dagger n+1,2} =\;& \frac{2\mathbf{P}\vec{\mathbf{u}} ^{\dagger n+1,2}-2\vec{\mathbf{u}} ^{\dagger n+1,4}}{\Delta t}  -  \vec{\mathbf{R}}_\mathrm{u}^{\dagger n+1,2}, \\
		\vec{\mathbf{N}}^{\dagger n+1,1} =\;& \frac{2\mathbf{P}\vec{\mathbf{u}} ^{\dagger n+1,1}-2\vec{\mathbf{u}} ^{\dagger n+1,4}}{\Delta t}  - 2\vec{\mathbf{R}}_\mathrm{u}^{\dagger n+1,1},\\
		\vec{\mathbf{N}}^{\dagger n,4} =\;& \frac{6\mathbf{P}\vec{\mathbf{u}} ^{\dagger n,4}-2(\vec{\mathbf{u}} ^{\dagger n+1,1}+2\vec{\mathbf{u}} ^{\dagger n+1,2}+\vec{\mathbf{u}} ^{\dagger n+1,3}-\vec{\mathbf{u}} ^{\dagger n+1,4})}{\Delta t} -  \vec{\mathbf{R}}_\mathrm{u}^{\dagger n,4},
	\end{align}
\end{subequations}
\begin{subequations}\label{eq:adj_vp}
	\begin{align}
		\vec{\mathbf{M}}^{\dagger n+1,3} =\;& \frac{\vec{\mathbf{v}}_\mathrm{p} ^{\dagger n+1,3}-\vec{\mathbf{v}}_\mathrm{p}^{\dagger n+1,4}}{\Delta t}  - \frac{1}{2}\vec{\mathbf{R}}_\mathrm{v_p}^{\dagger n+1,3},\\
		\vec{\mathbf{M}}^{\dagger n+1,2} =\;& \frac{2\vec{\mathbf{v}}_\mathrm{p} ^{\dagger n+1,2}-2\vec{\mathbf{v}}_\mathrm{p}^{\dagger n+1,4}}{\Delta t}  -  \vec{\mathbf{R}}_\mathrm{v_p}^{\dagger n+1,2}, \\
		\vec{\mathbf{M}}^{\dagger n+1,1} =\;& \frac{2\vec{\mathbf{v}}_\mathrm{p} ^{\dagger n+1,1}-2\vec{\mathbf{v}}_\mathrm{p}^{\dagger n+1,4}}{\Delta t}  -2\vec{\mathbf{R}}_\mathrm{v_p}^{\dagger n+1,1}, \\
		\vec{\mathbf{M}}^{\dagger n,4} =\;& \frac{6\vec{\mathbf{v}}_\mathrm{p} ^{\dagger n,4}-2(\vec{\mathbf{v}}_\mathrm{p}^{\dagger n+1,1}+2\vec{\mathbf{v}}_\mathrm{p} ^{\dagger n+1,2}+\vec{\mathbf{v}}_\mathrm{p} ^{\dagger n+1,3}-\vec{\mathbf{v}}_\mathrm{p} ^{\dagger n+1,4})}{\Delta t} - \vec{\mathbf{R}}_\mathrm{v_p}^{\dagger n,4},
	\end{align}
\end{subequations}
and
\begin{subequations}\label{eq:adj_xp}
	\begin{align}
		\vec{\mathbf{X}}^{\dagger n+1,3} =\;& \frac{\vec{\mathbf{x}}_\mathrm{p} ^{\dagger n+1,3}-\vec{\mathbf{x}}_\mathrm{p}^{\dagger n+1,4}}{\Delta t}  - \frac{1}{2}\vec{\mathbf{R}}_\mathrm{x_p}^{\dagger n+1,3}, \\
		\vec{\mathbf{X}}^{\dagger n+1,2} =\;& \frac{2\vec{\mathbf{x}}_\mathrm{p} ^{\dagger n+1,2}-2\vec{\mathbf{x}}_\mathrm{p}^{\dagger n+1,4}}{\Delta t}  -  \vec{\mathbf{R}}_\mathrm{x_p}^{\dagger n+1,2},\\
		\vec{\mathbf{X}}^{\dagger n+1,1} =\;& \frac{2\vec{\mathbf{x}}_\mathrm{p} ^{\dagger n+1,1}-2\vec{\mathbf{x}}_\mathrm{p}^{\dagger n+1,4}}{\Delta t}  -2\vec{\mathbf{R}}_\mathrm{x_p}^{\dagger n+1,1}, \\
		\vec{\mathbf{X}}^{\dagger n,4} =\;& \frac{6\vec{\mathbf{x}}_\mathrm{p} ^{\dagger n,4}-2(\vec{\mathbf{x}}_\mathrm{p}^{\dagger n+1,1}+2\vec{\mathbf{x}}_\mathrm{p} ^{\dagger n+1,2}+\vec{\mathbf{x}}_\mathrm{p} ^{\dagger n+1,3}-\vec{\mathbf{x}}_\mathrm{p} ^{\dagger n+1,4})}{\Delta t} - \vec{\mathbf{R}}_\mathrm{x_p}^{\dagger n,4},
	\end{align}
\end{subequations}
where the right-hand sides are defined as  
\begin{subequations}
	\begin{align}
		\vec{\mathbf{R}}_\mathrm{u}^{\dagger n,s} =\;& \mathbf{J}^T\Big[\mathbf{J}{(\vec{\mathbf{u}}^{n,s-1})}^T\odot\mathbf{G}{(\mathbf{P}\vec{\mathbf{u}}^{\dagger n,s})}^T\Big]+\mathbf{J}^T\Big[\mathbf{J}{(\vec{\mathbf{u}}^{n,s-1})}^T\odot\mathbf{G}\mathbf{P}\vec{\mathbf{u}}^{\dagger n,s}\Big]-\notag\\
		&{\Big( \upsilon L^2+\alpha L^{-4}+\nabla_{\vec{\mathbf{u}}^{n,s-1}} \vec{\mathbf{F}}_\mathrm{p}^{n,s-1} \Big)}^T\mathbf{P}\vec{\mathbf{u}}^{\dagger n,s} +{\Big(\nabla_{\vec{\mathbf{u}}^{n,s-1}}\vec{\mathbf{h}}_\mathrm{u}(\vec{\bm{\phi}}_u^{n,s-1};\vec{\theta}_{u})\Big)}^T\vec{\mathbf{u}}^{\dagger n,s} +\\
		&{\bigg( \frac{1}{\tau_p}\mathbf{B}+\nabla_{\vec{\mathbf{u}}^{n,s-1}}\vec{\mathbf{h}}_\mathrm{p}(\vec{\bm{\phi}}_p^{n,s-1};\vec{\theta}_{p}) \bigg)}^T\vec{\mathbf{v}}_\mathrm{p}^{\dagger n,s}+2\vec{K}^{n,s-1}_u\nabla_{\vec{\mathbf{u}}^{n,s-1}}\vec{K}^{n,s-1}_u,\notag\\
		\vec{\mathbf{R}}_\mathrm{v_p}^{\dagger n,s} =\;& \vec{\mathbf{x}}_\mathrm{p}^{\dagger n, s} +{\bigg(\nabla_{\vec{\mathbf{v}}_\mathrm{p}^{n,s-1}}\vec{\mathbf{h}}_\mathrm{p}(\vec{\bm{\phi}}_p^{n,s-1};\vec{\theta}_{p})-\frac{1}{\tau_p}\mathbf{I}\bigg)}^T\vec{\mathbf{v}}_\mathrm{p}^{\dagger n, s}-{\Big( \nabla_{\vec{\mathbf{v}}_\mathrm{p}^{n,s-1}}\vec{\mathbf{F}}_\mathrm{p}^{n,s-1}\Big)}^T\mathbf{P}\vec{\mathbf{u}}^{\dagger n,s}+\\
		& 2\vec{K}_{p}^{n,s-1}\nabla_{\vec{\mathbf{v}}_\mathrm{p}^{n,s-1}} \vec{K}_{p}^{n,s-1},\notag\\
		\vec{\mathbf{R}}_\mathrm{x_p}^{\dagger n,s} =\;& {\Big(\nabla_{\vec{\mathbf{x}}_\mathrm{p}^{n,s-1}}\vec{\mathbf{h}}_\mathrm{p}(\vec{\bm{\phi}}_p^{n,s-1};\vec{\theta}_{p})+\frac{1}{\tau_p}\nabla_{\vec{\mathbf{x}}_\mathrm{p}^{n,s-1}}\mathbf{B}\vec{\mathbf{u}}^{n,s-1} \Big)}^T\vec{\mathbf{v}}_\mathrm{p}^{\dagger n,s}- \\
		&{\Big(\nabla_{\vec{\mathbf{x}}_\mathrm{p}^{n,s-1}}\vec{\mathbf{F}}_\mathrm{p}^{n,s-1}\Big)}^T\mathbf{P}\vec{\mathbf{u}}^{\dagger n,s} + 2\vec{K}_{p}^{n,s-1}\nabla_{\vec{\mathbf{x}}_\mathrm{p}^{n,s-1}} \vec{K}_{p}^{n,s-1}.\notag
\end{align}
\end{subequations}

These equations hold for $n\in[N_t-1,0]$. 
The $N_t-1$ condition is obtained by gathering the terms for $n=N_t$ and $s=0$:
\begin{align*}
	 \vec{\mathbf{u}}^{\dagger N_t,4} &= \frac{\Delta t}{3}\vec{K}_{u}^{N_t,4}(\nabla_{\vec{\mathbf{u}}}\vec{K}_{u}^{N_t,4}),\\
	 \vec{\mathbf{v}}_\mathrm{p}^{\dagger N_t,4} &= \frac{\Delta t}{3}\vec{K}_{p}^{N_t,4}(\nabla_{\vec{\mathbf{v}}_\mathrm{p}}\vec{K}_{p}^{N_t,4}),\\
	 \vec{\mathbf{x}}_\mathrm{p}^{\dagger N_t,4} &= \frac{\Delta t}{3}\vec{K}_{p}^{N_t,4}(\nabla_{\vec{\mathbf{x}}_\mathrm{p}}\vec{K}_{p}^{N_t,4}).
\end{align*}
\section{Particle momentum correction}\label{app:hp_projection}
As discussed in section~\ref{sec:mom_cons_NN}, for momentum to be conserved in our closed flow system, 
\begin{equation}
	\sum_{j=1}^{N_p} \vec{\mathbf{h}}_{\mathrm{p}_j} = 0.
\end{equation}
The minimal adjustment for the neural-network closure $\vec{\mathbf{h}}_{\mathrm{p}_j}^\circ$, whose sum is nonzero, to conserve momentum is the orthogonal projection
\begin{equation}
	\vec{\mathbf{h}}_\mathrm{p} = \underset{\vec{\mathbf{h}}}{\mathrm{argmin}} \sum_{j=1}^{N_p}\| \vec{\mathbf{h}}_j -  \vec{\mathbf{h}}_{\mathrm{p}_j}^\circ \|^2 \quad \mathrm{s.t.}\quad \sum_{j=1}^{N_p} \vec{\mathbf{h}}_j = 0,
\end{equation}
where $\vec{\mathbf{h}}$ is the optimization variable.
The Lagrangian for this constrained optimization is 
\begin{equation}
	\mathcal{L} = \sum_{j=1}^{N_p}\| \vec{\mathbf{h}}_{\mathrm{p}_j} -  \vec{\mathbf{h}}_{\mathrm{p}_j}^\circ \|^2 + \lambda\sum_{j=1}^{N_p} \vec{\mathbf{h}}_{\mathrm{p}_j}  = 0,
\end{equation}
where $\| \cdot \|$ is the $L^2$ norm, and $\lambda$ enforces zero net-momentum addition.
Applying the first optimality condition
\begin{equation}\label{eq:hp_opt}
	\nabla_{\vec{\mathbf{h}}_{\mathrm{p}_j}}\mathcal{L} = 2(\vec{\mathbf{h}}_{\mathrm{p}_j}  -  \vec{\mathbf{h}}_{\mathrm{p}_j}^\circ) + \lambda = 0\Rightarrow \vec{\mathbf{h}}_{\mathrm{p}_j}  = \vec{\mathbf{h}}_{\mathrm{p}_j}^\circ - \frac{1}{2}\lambda
\end{equation}
and the constraint
\begin{equation}
	\sum_{j=1}^{N_p} \vec{\mathbf{h}}_{\mathrm{p}_j} = \sum_{j=1}^{N_p}(\vec{\mathbf{h}}_{\mathrm{p}_j}^\circ - \lambda) = 0,
\end{equation}
leads to 
\begin{equation}\label{eq:lmbda}
	\lambda = \frac{2}{N_p}\sum_{j=1}^{N_p}\vec{\mathbf{h}}_{\mathrm{p}_j}^\circ.
\end{equation}
Substitution of (\ref{eq:lmbda}) into (\ref{eq:hp_opt}) yields
\begin{equation}
	\vec{\mathbf{h}}_{\mathrm{p}_j}  = \vec{\mathbf{h}}_{\mathrm{p}_j}^\circ - \frac{1}{N_p}\sum_{j=1}^{N_p}\vec{\mathbf{h}}_{\mathrm{p}_j}^\circ,
\end{equation}
which is the momentum drift correction described in section~\ref{sec:mom_cons_NN}.

\bibliographystyle{unsrt_arxiv}
\bibliography{references}

\end{document}